\documentclass[journal]{IEEEtran}
\usepackage{cite}
\usepackage{amsmath,amssymb,amsfonts}
\usepackage{graphicx}
\usepackage{textcomp}
\usepackage{times}
\usepackage{amsmath}
\usepackage{amssymb}
\usepackage{epsfig}
\usepackage{epstopdf}
\usepackage{makecell}
\usepackage{tabularx}
\usepackage{booktabs}
\usepackage{multirow}
\usepackage{dcolumn}
\usepackage{threeparttable}
\usepackage{setspace}
\usepackage{algpseudocode}
\usepackage{algorithm}
\usepackage{color}
\usepackage{verbatim}
\usepackage{array}
\usepackage[table,xcdraw]{xcolor}
\usepackage{mathrsfs}
\usepackage{graphicx}
\usepackage{multirow}
\usepackage{adjustbox}
\usepackage{epsfig,epstopdf}
\usepackage{bbding}
\usepackage{subfigure}
\usepackage{adjustbox}

\usepackage[numbers, sort&compress]{natbib}

\newcommand{\widthscalefive}{0.148}
\newcommand{\widthscalefour}{0.120}

\usepackage[switch]{lineno}
\modulolinenumbers[5]

\newcommand{\RN}[1]{%
	\textup{\lowercase\expandafter{\it \romannumeral#1}}%
}

\def\BibTeX{{\rm B\kern-.05em{\sc i\kern-.025em b}\kern-.08em
    T\kern-.1667em\lower.7ex\hbox{E}\kern-.125emX}}

\ifCLASSINFOpdf
\else
\fi
\ifCLASSOPTIONcompsoc
 \usepackage[caption=false,font=normalsize,labelfont=sf,textfont=sf]{subfig}
\else
 \usepackage[caption=false,font=footnotesize]{subfig}
\fi
\begin{document}

%
\title{Fine-grained Attention and Feature-sharing Generative Adversarial Networks \\for Single Image Super-Resolution}
%
%
%

\author{Yitong~Yan,
        Chuangchuang~Liu,
        Changyou~Chen,
        Xianfang~Sun,
        Longcun~Jin*,~\IEEEmembership{Member,~IEEE,}
        Xinyi~Peng,
        and~Xiang~Zhou
\thanks{Y. Yan and C. Liu contribute equally in this work and share the first authorship.}
\thanks{*Corresponding author: Longcun Jin (lcjin@scut.edu.cn).}
\thanks{Y. Yan, C. Liu, L. J and X. Peng were with the School of Software Engineering, South China University of Technology, Guangzhou, China (e-mail: seyannis\_yan@mail.scut.edu.cn; selcc@mail.scut.edu.cn; lcjin@scut.edu.cn; adxypeng@scut.edu.cn).}
\thanks{C. Chen was with the Department of Computer Science and Engineering, University at Buffalo, State University of New York, NY, USA (e-mail: changyou@buffalo.edu).}
\thanks{X. Sun was with the School of Computer Science and Informatics Cardiff University, UK (e-mail: sunx2@cardiff.ac.uk).}
\thanks{X. Zhou was with the School of Data Science and Department of Mathematics at College of Science, City University of Hong Kong, Hong Kong, China (e-mail: Xiang.Zhou@cityu.edu.hk).}
}

%
%

\markboth{Journal of \LaTeX\ Class Files,~Vol.~14, No.~8, August~2015}%
{Shell \MakeLowercase{\textit{et al.}}: Bare Demo of IEEEtran.cls for IEEE Journals}
%



\maketitle

\begin{abstract}
Traditional super-resolution (SR) methods by minimize the mean square error usually produce images with over-smoothed and blurry edges, due to the lack of high-frequency details. In this paper, we propose two novel techniques within the generative adversarial network framework to encourage generation of photo-realistic images for image super-resolution. Firstly, instead of producing a single score to discriminate real and fake images, we propose a variant, called Fine-grained Attention Generative Adversarial Network (FASRGAN), to discriminate each pixel of real and fake images. FASRGAN adopts a UNet-like network as the discriminator with two outputs: an image score and an image score map. The score map has the same spatial size as the HR/SR images, serving as the fine-grained attention to represent the degree of reconstruction difficulty for each pixel. 
Secondly, instead of using different networks for the generator and the discriminator, we introduce a feature-sharing variant (denoted as Fs-SRGAN) for both the generator and the discriminator. The sharing mechanism can maintain model express power while making the model more compact, and thus can improve the ability of producing high-quality images. Quantitative and visual comparisons with state-of-the-art methods on benchmark datasets demonstrate the superiority of our methods. We further apply our super-resolution images for object recognition, which further demonstrates the effectiveness of our proposed method. 
The code is available at \underline{https://github.com/Rainyfish/FASRGAN-and-Fs-SRGAN}.
\end{abstract}


%
\IEEEpeerreviewmaketitle

\section{Introduction}
%
%
%
%
\IEEEPARstart{S}{ingle} image super-resolution (SISR), which aims to recover a high-resolution (HR) image from its low-solution (LR) version, has been an active research topic in computer graphic and vision for decades. SISR has also attracted increasing attention in both academia and industry, with applications in various fields such as medical imaging, security surveillance, 
object recognition and so on. However, SISR is a typically ill-posed problem due to the irreversible image degradation process, {\it i.e.}, multiple HR images can be generated from one single LR image. Learning the mapping between HR and LR images plays an important role in addressing this problem.

Recently, deep convolution neural networks (CNNs) have been shown great success in many vision tasks, such as image classification, object detection, and image restoration. Dong \emph{et al.}~\cite{dong2015image} first proposed a three-layer CNN for single image super-resolution (SRCNN) to directly learn the 
mapping from LR to HR images. Since then the CNN-based methods~\cite{9044873} have been dominant for the SR problem because they greatly improved the reconstruction performance. Kumar \emph{et al.}~\cite{kumar2016fast} tapped into the ability of polynomial neural networks to hierarchically learn refinements of a function that maps LR to HR patches. VDSR~\cite{kim2016accurate} obtained remarkable performance by increasing the depth of the network to 20, proving the importance of the network depth for detecting effective features of images. 
EDSR~\cite{lim2017enhanced} removed unnecessary batch normalization layer in the ResNet~\cite{he2016deep} architecture and widened the channels, significantly improving the performance. 
RCAN~\cite{zhang2018image} applied residual in residual structure to construct a very deep network and used a channel attention mechanism to adaptively rescale features. 

\begin{figure*}
    \centering
    \includegraphics[width=0.7\linewidth]{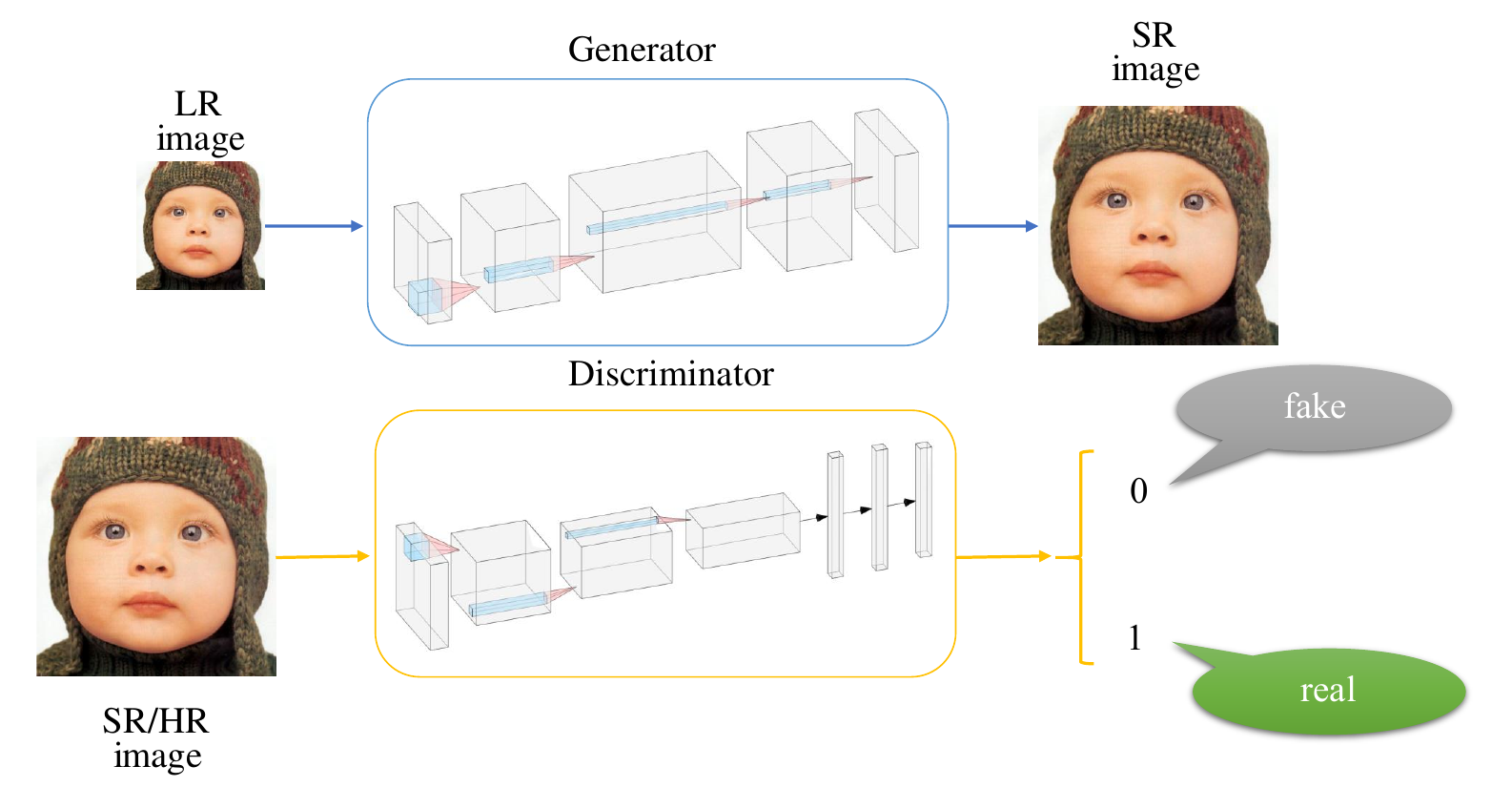}
    \caption{The architecture of GAN-based Super-Resolution method. The generator aims to reconstruct photo-realistic SR images, while the discriminator
    distinguishes the SR image from the ground-truth HR image.}
    \label{fig:GAN_based}
\end{figure*}

The aforementioned methods use the optimization idea of minimizing the mean squared error (MSE) between the recovered SR image and the corresponding HR image. Such methods are designed to maximize the peak signal-to-noise ratio (PSNR). However, they typically produce over-smoothed edges and lack tiny textures. To produce photo-realistic SR images, Ledig \emph{et al.}\cite{ledig2017photo} first used 
the generative adversarial network (GAN)~\cite{goodfellow2014generative} to match the underlying distributions of HR and SR images. 
ESRGAN~\cite{wang2018esrgan} further extended the generator network and used the Relativistic Discriminator~\cite{jolicoeur-martineau2018} to produce more photo-realistic images. 
However, as shown in Fig.\ref{fig:GAN_based}, the discriminator in these GAN-based methods only outputs a score of the whole input SR/HR image, which is a coarse way to guide the generator. Furthermore, the generator and discriminator of these works are considered to be independent, while we believe there should be significant information to be shared. For example, the lower-level parts of the two networks both aim at extracting tiny features such as corners and edges, which we believe should be correlated. 

\begin{figure*}
    \centering
    \includegraphics[width=0.7\linewidth]{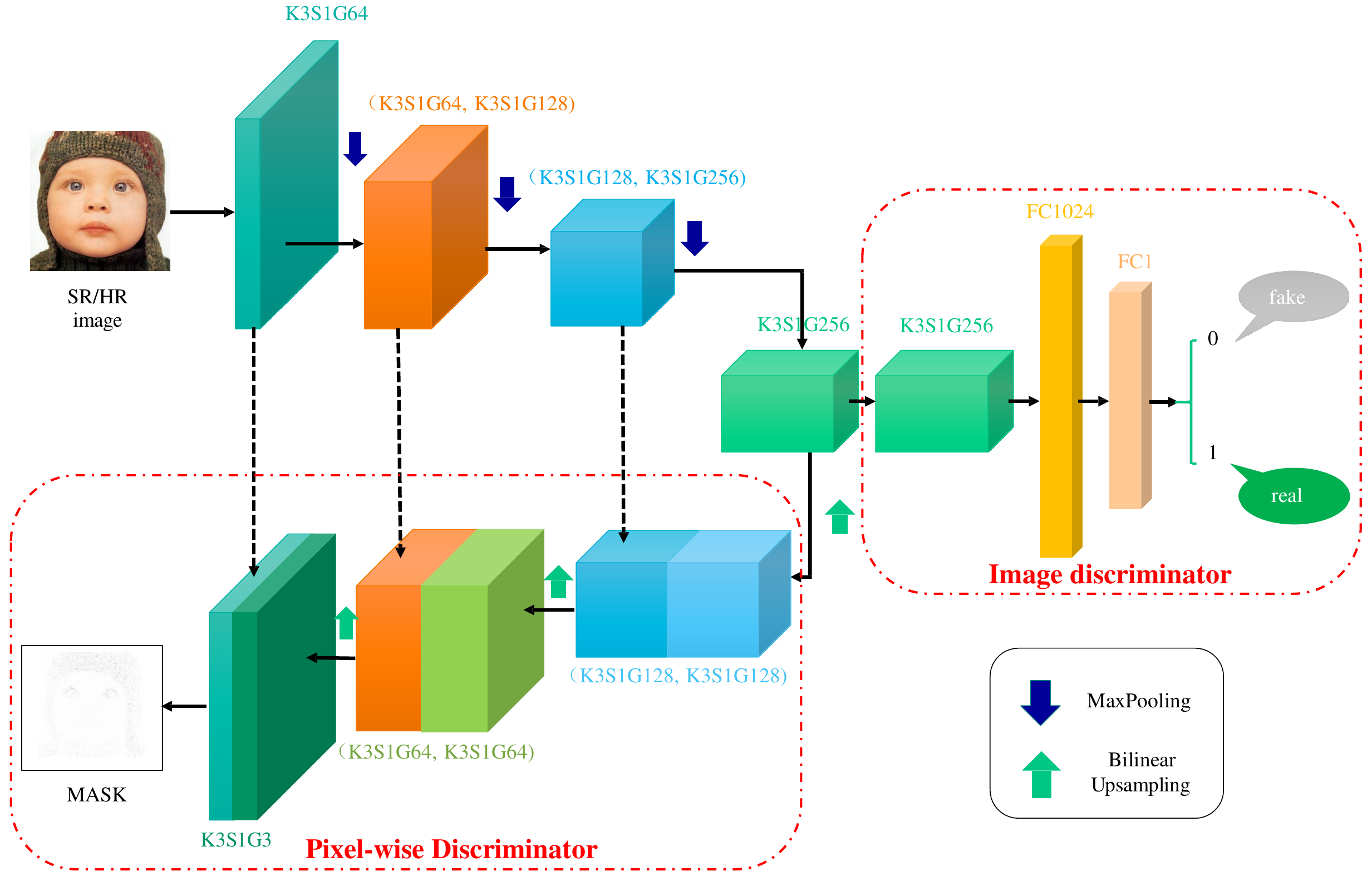}
    \caption{The discriminator architecture of FASRGAN, where K, S, G represent the kernel size, the stride, and the filter number of the Conv layer, respectively. FC stands for fully connected layer. The mask is a score map among $[0, 1]$, donating the difficulty of reconstruction of each pixel in the image. } 
    \label{fig:FASRGAN_D}
\end{figure*}

To address these limitations, we propose two novel techniques based on the GAN framework for image super-resolution, a fine-grained attention mechanism for the discriminator and a feature-sharing network component for both the generator and the discriminator. Specifically, we use a UNet-like~\cite{ronneberger2015u} discriminator (Fig.\ref{fig:FASRGAN_D}) to introduce a fine-grained attention in the GAN (FASRGAN). Our discriminator produces two outputs, a score of the whole input image and a fine-grained score map of every pixel in the image. The score map shares the same spatial size as the input image, and measures the degree of differences at each pixel between the generated and the true distributions. To produce better visual quality images, we incorporate the score map into the loss function with an attention mechanism to make the generator pay more attention on the hard parts of an image, instead of treating all parts equally. In addition, we propose a feature-sharing mechanism (Fig.\ref{fig:Co-SRGAN}) to align the low-level feature extraction of both the generator and the discriminator (Fs-SRGAN). This novel structure can significantly reduce the number of parameters and improve the performance. 

Overall, our main contributions are three-fold:
\begin{itemize}
    \item We propose a novel UNet-like discriminator to generate a single score for the whole image and a pixel-wise score map of the input image. We further incorporate the score map into the loss function with an attention mechanism to define the generator. This attention mechanism makes the generator focus on hard parts of an image for generation. 
    \item We introduce a feature-sharing mechanism to define the shared low-level feature extraction for the generator and the discriminator. This reduces the number of model parameters and helps the generator and the discriminator extract more effective features.
    \item The proposed two components are general, and can be applied to other GAN-based SR models. 
    Extensive experiments on benchmark datasets illustrate the superiority of our proposed methods compared with current state-of-the-art methods.
\end{itemize}

The remainder of the paper is organized as follows. Section II describes related works. The proposed GAN-based methods are presented in Section III. Experimental results are discussed in Section IV. Finally, the conclusions are drawn in Section V.

\section{Related Work}
Traditional SISR methods are exemplar or dictionary based. However, these methods are limited by the size of datasets or dictionaries, 
and are usually time-consuming. These shortcomings can be greatly alleviated by the recent CNN-based methods~\cite{9044873}.

In their pioneer work, Dong \emph{et al.}~\cite{dong2015image} applied convolutional neural networks with three layers for SISR, namely SRCNN, to learn a mapping from LR to HR images in an end-to-end manner. Kim \emph{et al.}~\cite{kim2016accurate} increases the depth of the network to 20, achieving great improvement in accuracy compared to SRCNN. 
Instead of using the interpolated LR images as the inputs of network, FSRCNN~\cite{dong2016accelerating} extracted features from the origin LR images and upscaled the spatial size by upsampling layers at the tail of the network. This architecture is widely used in the subsequent SR methods. Various advanced upsampling structures have been proposed recently, for instance, deconvolutional layer~\cite{zeiler2010deconvolutional} 
, sub-pixel convolution~\cite{shi2016real}, and EUSR~\cite{kim2018deep}. LapSRN~\cite{lai2017deep} 
progressively reconstructed an HR image with increasing scales of an input image by the Laplacian pyramid structure. 
Lim \emph{et al.}~\cite{lim2017enhanced} proposed a very large network (EDSR) and its multi-scale version (MDSR), which removed the unnecessary batch normalization layer in the ResNet~\cite{he2016deep} and greatly improved super-resolution performance. D-DBPN~\cite{haris2018deep} introduced an error-correcting feedback mechanism to learn relationships between LR features and SR features.
ZSSR~\cite{shocher2018zero} used a unsupervised method to learn the mapping between HR images and LR images. DIP~\cite{ulyanov2018deep} showed that the structure of a generator network can capture a large amount of low-level image statistics before any learning is performed, 
which can be used as a handcrafted prior with excellent results in super-resolution and other standard inverse problems. To address the real-world LR image problem, Fritsche \emph{et al.}~\cite{fritsche2019frequency} proposed to separate the low and high image frequencies and treat them in different ways during training. Adversarial training is used to modify only the high, not the low frequencies. 
RDN~\cite{zhang2018residual} combined dense and residual connections to make full use of information of LR images. Different from RDN, MS-RHDN \cite{liu2019multi} proposed multi-scale residual hierarchical dense networks to extract multi-scale and hierarchical feature maps.
Yang \emph{et al.}~\cite{yang2018drfn} proposed a deep recurrent fusion network (DRFN) for SR with large-scale factors, which used transposed convolution to jointly extract and upsample raw features from the input and used multi-level fusion for reconstruction. SeaNet~\cite{fang2020soft} proposed a Soft-edge assisted Network to reconstruct the high-quality SR images with
the help of image soft-edge. Zhang \emph{et al.}~\cite{zhang2020adaptive} proposed an adaptive importance learning scheme to enhance the performance of the lightweight SISR network architecture.
RCAN \cite{zhang2018image} applied channel-attention mechanism to adaptively rescale  channel-wise features. SAN \cite{Dai_2019_CVPR} further  proposed a second-order channel attention (SOCA) module to rescale the features 
instead of global average pooling. 

The aforementioned methods aim to achieve high PSNR and SSIM~\cite{wang2004image} values. However, these criteria usually cause heavy over-smoothed edges and artifacts. Images generated by these MSE-based SR methods lose various high-frequency details and have a bad perceptual quality. To generate more photo-realistic images, Ledig \emph{et al.}~\cite{ledig2017photo} firstly introduced generative adversarial network into image super-resolution, called SRGAN. SRGAN combined a perceptual loss and an adversarial loss to improve the reality of generated images. But some visually implausible artifacts still could be found in some generated images. To reduce the artifacts, EnhanceNet~\cite{sajjadi2017enhancenet} combined a pixel-wise loss in the image space, a perceptual loss in the feature space, a texture matching loss 
and an adversarial loss. 
The contextual loss~\cite{mechrez2018maintaining} was a kind of perceptual loss to make the generated images as similar as possible to ground-truth images. 
Yan \emph{et al.}~\cite{yan2019deep} 
firstly trained a novel full-reference image quality assessment (FR-IQA) approach for SISR, then employed the proposed loss function (SR-IQA) to train their SR network which contains their proposed highway unit. In addition, they also integrated SR-IQA loss to the GAN-based SR method to achieve better results for both accuracy and perceptual quality.
Based on SRGAN, ESRGAN~\cite{wang2018esrgan} $\RN{1}$) substituted the standard residual block with a residual-in-residual dense block, $\RN{2})$ removed batch normalization layers, $\RN{3})$ utilized VGG feature before activated as perceptual loss, and $\RN{4})$ replaced the standard discriminator with Relativistic Discriminator proposed in RaGAN~\cite{jolicoeur-martineau2018}. 
Noteworthily, ESRGAN won the first place in the 2018 PIRM Challenge on Perceptual Image Super-Resolution~\cite{blau20182018}, which pursued the high perceptual-quality images. RankSRGAN~\cite{zhang2019ranksrgan} firstly trained a ranker to learn the behavior of perceptual metrics and then introduced a rank-content loss to optimize the perceptual quality.

\section{Proposed Methods}
\subsection{Overview}

Our methods aim to reconstruct a high-resolution image $I^{\text{SR}}\in R^{Wr\times Hr\times C}$ from a low-resolution image $I^{\text{LR}}\in R^{W\times H\times C}$, where $W$ and $H$ are the width and height of the LR image,  $r$ is the upscaling factor, and $C$ is the number of channels of the color space.
This section details our two strategies within the GAN framework for image super-resolution in order: FASRGAN and Fs-SRGAN. Specifically, we propose a fine-grained attention mechanism in FASRGAN to make the generator focus on the difficult parts of image reconstruction based on the score map from the UNet-like discriminator, instead of treating each part equally. We further propose a feature-sharing mechanism in Fs-SRGAN by sharing the low-level feature extractor of the generator and the discriminator. Both networks update the gradient of the shared low-level feature 
extractor in the training phase, which could make the model more compact while keeping enough representation power. 
These two strategies contribute to the overall perceptual quality for SR, respectively. For simplicity, we use the same network architecture  as ESRGAN~\cite{wang2018esrgan} for the generator to generate the SR images from the input LR images.

\subsection{Fine-grained Attention Generator Adversarial Networks}
Our proposed fine-grained attention GAN (FASRGAN) designs a specific discriminator for SISR. As discussed above and shown in Fig.\ref{fig:GAN_based}, the discriminator in a standard GAN-based SR model outputs a score of the whole input SR/HR image. This can be considered as a coarse way to judge an input image and cannot discriminate local features of inputs. To tackle this problem, the proposed FASRGAN defines a UNet-like discriminator contained two outputs, corresponding to a score of the whole image and a fine-grained score map. The score map has the same size as the input image and is used for pixel-wise discrimination. The proposed discriminator is illustrated in Fig.~\ref{fig:FASRGAN_D}.

\subsubsection{\textbf{A UNet-like Discriminator}}
The UNet-like discriminator with two outputs can be divided into two parts: an encoder and a decoder. 

\textbf{Encoder. } 
Similar to the standard discriminator D in ESRGAN, the encoder part of the proposed UNet-like discriminator uses a standard max-pooling layer with a stride of 2 to reduce the spatial size of a feature map and increase receptive fields. At the same time, the number of channels is increased for improving representative ability. At the end of the encoder, two fully connected layers are added to output a score, measuring the overall probability of an input image $x$ being real or fake. We further enhance the discriminator based on the Relativistic GAN~\cite{jolicoeur-martineau2018}, which has also been used in ESRGAN~\cite{wang2018esrgan}. The loss function is defined as: 
\begin{equation}\label{eq:discriminator_adversarial}
\begin{split}
    L_{adv}^D = & \mathbb{E}_{x_r}[\log(1-D_{Ra}(x_r,x_f))] \\
               & + \mathbb{E}_{x_f}[\log(D_{Ra}(x_f,x_r))] \\
             = &\mathbb{E}_{x_r}[\log(1-\sigma(C(x_r)-\mathbb{E}_{x_f}[C(x_f)]))] \\
               & + \mathbb{E}_{x_f}[\log(\sigma(C(x_f)-\mathbb{E}_{x_r}[C(x_r)]))],
\end{split}
\end{equation}
where $x_r$ and $x_f$ stand for the ground-truth image and the generated SR image, respectively. $D_{Ra}(\cdot)$ refers to the function of the relativistic discriminator, which tries to predict the probability that a real image $x_r$ is more realistic than a fake one $x_f$; $C(x)$ is the 
discriminator output before sigmoid function and $\sigma$ is the sigmoid function.

\textbf{Decoder. } We exploit an upsampling layer to gradually extend the spatial size of feature maps, followed by two convolutional layers to extract more information. To make full use of features, we concatenate the previous outputs from the encoder, which have the same spatial size as current ones. As shown in Fig.~\ref{fig:FASRGAN_D}, the feature maps at the end of the decoder have the same spatial size as input images. Finally, we use the sigmoid function to produce a score map $M\in R^{Wr\times Hr \times C}$ that provides pixel-wise discrimination between real and fake pixels of an input image. The fine-grained adversarial loss function $L_{M}^D$ for the discriminator is defined as:
\begin{align}\label{eq:mask_d}
     & L_{M}^D = \frac{1}{Wr\times Hr\times C}\\
     & \times \sum_{c=1}^C\sum_{w=1}^{Wr}\sum_{h=1}^{Hr}\{{\log(1-M_r(w,h,c))}+{\log(M_f(w,h,c))}\}, \nonumber
\end{align}
where $M_r$ and $M_f$ represent the score maps of the HR image and the generated SR image, respectively. 
 Finally, the loss function for the discriminator is defined as: $L^D = L^D_{adv} + L^D_M$.
\begin{figure*}
    \centering
    \includegraphics[width=0.7\linewidth]{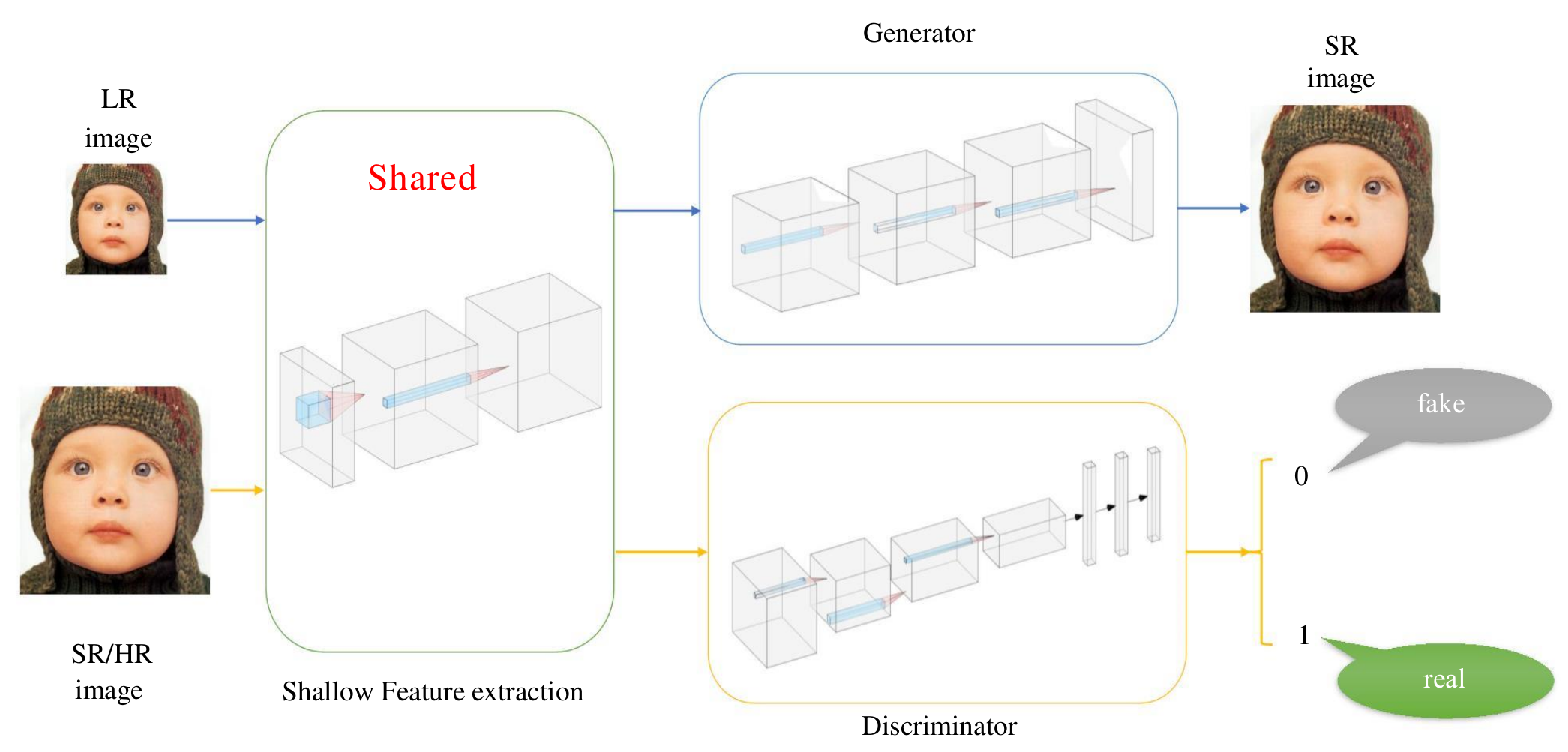}
    \caption{The illustration of our Feature-sharing Generator Adversarial Networks (Fs-SRGAN). The input sizes of the generator and the discriminator are different. We use a fully Convolutional Neural Network with invariant size of the feature map so that the different input sizes do not matter. }
    \label{fig:Co-SRGAN}
\end{figure*}

 \subsubsection{Generator Objective Function}
In the GAN-based SR methods, the generator is generally used to generate the SR images from the LR images. ESRGAN~\cite{wang2018esrgan} introduced Residual-in-Residual Dense Block (RRDB) without batch normalization as the basic network building unit, which is of higher capacity and easier to train compared with the ResBlock in SRGAN~\cite{ledig2017photo}. In this paper, we also adopt RRDB to construct our generator for a fair comparison with ESRGAN. 
The generator is trained by several losses, defined as following: 

\textbf{Content Loss. } Following~\cite{lai2017deep, MSLapSRN, lim2017enhanced, zhang2018residual}, we use an $L_1$ loss function to constrain the content of a generated SR image to be close to the HR image. The loss is defined in Eq.\ref{eq:L1}. 
\begin{align}\label{eq:L1}
    &L_1=\frac{1}{Wr\times Hr \times C} \\                
    &\times \sum_{w=1}^{Wr}\sum_{h=1}^{Hr}\sum_{c=1}^C  \parallel F_{\theta}^{G} (I_i^{LR})(w,h,c)-I_i^{HR}(w,h,c)\parallel _1, \nonumber
\end{align}
where $F_\theta^G(\cdot)$ represents the function of the generator, $\theta$ is the parameters of the generator and $I_i$ means  the $i$-th image.

\textbf{Perceptual Loss. } 
The perceptual loss~\cite{johnson2016perceptual} aims to make the SR image close to the corresponding HR image based on high-level features extracted from a pre-trained network. Similar to \cite{ledig2017photo, wang2018esrgan}, we consider both the SR and HR images as the input to the pre-trained VGG19 and extract the VGG19-54 layer features. The perceptual loss is defined as:
\begin{equation}\label{eq:percep}
    L_{percep} = \parallel F_{\theta}^{VGG} (G(I_i^{LR}))-F_{\theta}^{VGG} (I_i^{HR})\parallel _1,
\end{equation}
where $F_{\theta}^{VGG}(\cdot)$ is the function of VGG and $I_i$ is the $i$-th image, $G(\cdot$) is the function of the generator.

\textbf{Adversarial Loss. } 
The discriminator contains two outputs, a whole estimation of the entire image and the pixel-wise fine-grained estimations of an input image. The total adversarial loss function for the generator is defined as:
\begin{equation} \label{eq:adv}
    L_{adv}^G = L_{entire}^G+L_{fine}^G,
\end{equation}
As shown in Eq.\ref{eq:mask_d}, the discriminator tries to distinguish the real and fake image in a fine-grained way, while the generator aims to fool the discriminator. Thus the fine-grained adversarial loss function $L^G_{fine}$ for the generator is the symmetrical form of Eq.\ref{eq:mask_d}:
 \begin{equation}
 \begin{split}\label{eq:mask_g}
      & L_{fine}^G = \frac{1}{Wr\times Hr\times C}\\
      & \times \sum_{c=1}^C\sum_{w=1}^{Wr}\sum_{h=1}^{Hr}\{{\log(M_r(w,h,c))}+{\log(1-M_f(w,h,c))}\},
 \end{split}
 \end{equation}
$L_{entire}^G$ is also the symmetrical form of Eq.\ref{eq:discriminator_adversarial} and defined as:
\begin{equation} \label{eq:entire_adv}
\begin{split}
    L_{entire}^G=&\mathbb{E}[\log(\sigma(C(x_r)-\mathbb{E}[C(x_f)]))] \\
            &+\mathbb{E}[\log(1-\sigma(C(x_f)-\mathbb{E}[C(x_r)]))].
\end{split}
\end{equation}

\textbf{Fine-grained Attention Loss Function.} 
The score map generated by the UNet-like discriminator is represented as pixel-wise discrimination scores of an input image, with values $M(w,h,c)$ among $[0,1]$. 
A higher score means the corresponding pixel of the input image is closer to that of the ground-truth image. In this manner, the score map can indicate which parts of an image are more difficult to generate and which parts are easier. For instance, the structure background part of an image is sometimes simpler, and thus it would expect the discriminator reflects this to the generator when updating the generator.  
In other words, the part with lower scores (more difficult to generate) should receive more attention when updating the generator. As a result, we incorporate the score map as the fine-grained attention mechanism into a $L_1$ loss function, constituting a weighted attention loss function:

\begin{equation}
    \begin{split}
    L_{attention}=&\frac{1}{Wr\times Hr \times C}\sum_{w=1}^{Wr}\sum_{h=1}^{Hr}\sum_{c=1}^C (1-M_f(w,h,c))\\
                  & \times \parallel F_{\theta}^{G} (I_i^{LR})(w,h,c)-I_i^{HR}(w,h,c)\parallel _1,
    \end{split}
    \label{eq:atten}
\end{equation}
where $M_f(w,h,c)$ is the score map of the generated image given by the discriminator. Instead of treating every pixel of an image equally, $L_{attention}$ contributes to pay more attention in the hard-to-recovered part of an image, such as the textures with rich semantic information.

Combining the above losses with different weights, the total loss of the generator is:
\begin{equation}\label{eq:generator_loss}
    L^{G}=  L_{percep} + \lambda_1 L_{adv}^G + \lambda_2 L_{attention} + \lambda_3 L_1,
\end{equation}
where $\lambda_1$, $\lambda_2$, $\lambda_3$ 
are the coefficients to balance different loss terms.

\subsection{Feature-sharing Generator Adversarial Networks}
In the standard GANs, the generator and the discriminator are usually defined as two independent networks. Based on the observation that the shallow parts of the two networks always extract 
low-level textures such as edges and corners, we propose a new network structure (Fs-SRGAN) to enable low-level feature sharing between the generator and the discriminator. This can reduce the number of parameters and help both networks extract more effective features. Consequently, our Fs-SRGAN contains three parts: a shared feature extractor, a generator, and a discriminator, as shown in Fig.~\ref{fig:Co-SRGAN}.

\subsubsection{\textbf{Shared Feature Extractor}}
 We first use a share feature extractor to transform an input image from color space to feature space, before extracting low-level feature maps. The feature-sharing mechanism allows the generator and the discriminator to jointly optimize the low-level feature extractor. 
Similar to FASRGAN, we adopt RRDB, the basic block of ESRGAN~\cite{wang2018esrgan}, 
as the basic structure. The shared feature extractor contains $E$ 
RRDBs to extract helpful feature maps for both the generator and the discriminator, described as following:
\begin{equation}\label{eq:shared_part}
    H_{shared} = F_{shared}(x),
\end{equation}
where $H_{shared}$ is the low-level feature maps extracted by the shared part, $F_{shared}$ represents the function of the shared feature extractor, and $x$ is the input. For the generator, the input is an LR image, while for the discriminator it is a SR image or a HR image. Considering the input sizes of the generator and the discriminator are different, we apply a fully Convolutional Neural Network with invariant size of feature maps to extract features so that the different input sizes do not matter.



\begin{figure*}[]
	\newlength\fsdttwofigBD
	\setlength{\fsdttwofigBD}{-0.4cm}
	\newlength\fsdbigfig
	\setlength{\fsdbigfig}{-0.38cm}
	\scriptsize
	\centering
	
	\begin{tabular}{c}	
    	\hspace{-2mm}
		\begin{adjustbox}{valign=t}
		\tiny
			\begin{tabular}{ccccc}
			   \includegraphics[width=0.148\textwidth]{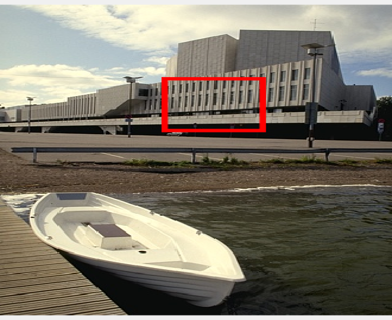} \hspace{\fsdbigfig}&
				\includegraphics[width=\widthscalefour \textwidth]{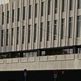} \hspace{\fsdttwofigBD} &
				\includegraphics[width=\widthscalefour \textwidth]{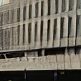} \hspace{\fsdttwofigBD} &
				\includegraphics[width=\widthscalefour \textwidth]{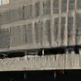} \hspace{\fsdttwofigBD} &
				\includegraphics[width=\widthscalefour \textwidth]{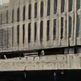} \hspace{\fsdttwofigBD}
				\\
				78004 from BSD100 ($4\times$):  \hspace{\fsdbigfig} &
				HR \hspace{\fsdttwofigBD} &
    			ESRGAN ~\cite{wang2018esrgan}\hspace{\fsdttwofigBD} &
				RankSRGAN~\cite{zhang2019ranksrgan}\hspace{\fsdttwofigBD} &
				FASRGAN (ours) \hspace{\fsdttwofigBD}
				\\
				PSNR/PI/LPIPS \hspace{\fsdbigfig} &
				$\infty$/2.5379/0 \hspace{\fsdttwofigBD} &
				24.90/2.4666/0.0577 \hspace{\fsdttwofigBD} &
				25.49/1.8918/0.0724 \hspace{\fsdttwofigBD} &
				25.57/2.2652/0.0537 \hspace{\fsdttwofigBD}

				\\
			\end{tabular}
		\end{adjustbox}
    \vspace{0.5mm}

 	\end{tabular}
	\caption{
		A visual comparison between the state-of-the-art perceptual image SR methods and FASRGAN (ours) for $4\times$ SR.
		}
	\label{fig:78004}
\end{figure*}

\subsubsection{\textbf{The Generator and the Discriminator}}
The rest parts of the generator and the discriminator are the same as those in standard GAN-based methods, except that the inputs are feature maps instead of images as shown in Fig.\ref{fig:Co-SRGAN}.

\textbf{Generator. } 
The generator contains three parts: low-level feature extraction, deep feature extraction, and reconstruction, which are used for transforming the input image to the feature space from the color space and extracting low-level information, extracting high-level semantic features and reconstructing SR image, respectively. Note that the generator in our Fs-SRGAN only contains the latter two parts. Similar to the shared low-level feature extraction, we adopt RRDB as the basic part of deep feature extraction, except that more RRDBs are used to increase the depth of the network with the purpose of extracting more high-frequency feature for reconstruction. The reconstruction part utilizes an upsampling layer to upscale the high-level feature maps and a Conv layer to reconstruct an SR image. The loss function of the generator is same as that of ESRGAN~\cite{wang2018esrgan}, which includes perceptual loss, adversarial loss, and pixel-based loss:
\begin{equation}\label{eq:fs_generator}
    L^{G}=  L_{percep} + \lambda_1 L_{adv}^G + \lambda_2 L_1,
\end{equation}
where $\lambda_1$, $\lambda_2$ are the coefficients to balance different loss terms, $L_{percep}$ and $L_1$ are defined in Eq.\ref{eq:percep} and Eq.\ref{eq:L1}, respectively, $L_{adv}^G$ is the adversarial loss with the same definition as $L^G_{entire}$ in Eq.\ref{eq:entire_adv}.

\textbf{Discriminator. } 
Because the discriminator is a classification network that distinguishes the input as SR or HR image, we apply a structure similar to the VGG network as the discriminator. To reduce information loss, we substitute the pooling layer (used in the encoder of the UNet-discriminator) for a Conv layer with a stride of 2 to decrease the size of feature map. At the tail of the discriminator, we use a Conv layer to transform the feature map into a one-dimensional vector, then use two fully connected layers to output the classification score $s$ among $[0, 1]$. The value of $s$ closer to 1 means more real, otherwise more fake. The loss function of the discriminator is same as $L^D_{adv}$ defined in Eq.\ref{eq:discriminator_adversarial}.

\section{Experimental Results}
In this section, we first describe our model training details, then provide quantitative and visual comparisons with several state-of-the-art methods on benchmark datasets for our two proposed novel methods, FASRGAN and Fs-SRGAN. 
We further combine the fine-grained attention and the feature-sharing mechanisms into one single model, termed FA$+$Fs-SRGAN.

\begin{table*}[]
\footnotesize

\center
\begin{center}
\caption{Quantitative results with the Bicubic degradation model for $4\times$ SR. Best and second best results are \textbf{highlighted} and \underline{underlined}, respectively.}
\label{tab:quantitative}
\begin{tabular}{c|c|c|c|c|c|c|c|c}
\hline
\multirow{2}{*}{Dataset} & \multirow{2}{*}{Metric} & {EnhanceNet} &  {SRGAN} &  {ESRGAN} & {RankSRGAN} &  {FASRGAN} & {Fs-SRGAN} & {FA+Fs-SRGAN}
\\
 & & \cite{sajjadi2017enhancenet}
 &\cite{ledig2017photo}
 & \cite{wang2018esrgan}
 & \cite{zhang2019ranksrgan}
 &(ours)
 &(ours)
 &(ours)
\\
\hline
\multirow{4}{*}{Set5} 
& PSNR & 28.57 & 29.91 & \textbf{30.46} & 29.73 & 30.15 & \underline{30.28} & 30.19 
\\ & SSIM & 0.8102 & 0.8510 & 0.8515 & 0.8398 & 0.8450 & \textbf{0.8588} & \underline{0.8571}
\\ & PI & \textbf{2.8466} & 3.4322 & 3.5463 & \underline{2.9867} & 3.1685 & 3.9143 & 3.7455
\\ & LPIPS & 0.0488 & 0.0389 & 0.0350 & 0.0348 &\textbf{0.0325} & \underline{0.0330} & 0.0344
\\
\hline
\multirow{4}{*}{Urban100} 
& PSNR & 23.54 & 24.39 & 24.36 & 24.49 & 24.51 & \underline{24.55} & \textbf{24.67}
\\ & SSIM & 0.6933 & 0.7309 & 0.7341 & 0.7319 & 0.7380 & \textbf{0.7509} &\underline{0.7466}
\\ & PI & \underline{3.4543} & 3.4814 & 3.7312 & \textbf{3.3253} & 3.5173 & 3.5940 & 3.5819 
\\ & LPIPS &0.0777 & 0.0693 & \underline{0.0591} & 0.0667 & \textbf{0.0588} & \underline{0.0591} & 0.0625 
\\
\hline
\multirow{4}{*}{BSD100} 
& PSNR & 24.94 & 25.50 & 25.32 & 25.51 & 25.41 & \underline{25.61} & \textbf{25.87}
\\ & SSIM & 0.6266 & 0.6528 & 0.6514 & 0.6530 & 0.6523 & \underline{0.6726} & \textbf{0.6747}
\\ & PI & 2.8467 & 2.3054 & 2.4150 & \textbf{2.0768} & \underline{2.2783} & 2.4056 & 2.3749 
\\ & LPIPS & 0.0982 & 0.0887 & \underline{0.0798} & 0.0850 & \textbf{0.0796} & 0.0801 & 0.0855
\\
\hline
\multirow{4}{*}{DIV2K val} 
& PSNR & 27.28 & 28.16 & \underline{28.17} & 28.10 & 28.15 & 28.15 & \textbf{28.23} 
\\ & SSIM & 0.7460 & 0.7753 & 0.7759 & 0.7710 & 0.7768 & \textbf{0.7903} & \underline{0.7891} 
\\ & PI & 3.4953 & 3.1619 & 3.2572 & \textbf{3.0130} & \underline{3.1034} & 3.3303 & 3.3092 
\\ & LPIPS & 0.0753 & 0.0605 & 0.0550 & 0.0576 & \textbf{0.0539} & \underline{0.0542} & 0.0576 
\\
\hline
\multirow{4}{*}{PIRM val} 
& PSNR & 25.47 & 25.61 & 25.18 & 25.65 & 25.38 & \underline{25.75} & \textbf{26.00} 
\\ & SSIM & 0.6569 & 0.6757 & 0.6596 & 0.6726 & 0.6648 & \underline{0.6907} & \textbf{0.6934} 
\\ & PI & 2.6762 & \underline{2.2254} & 2.5548 & \textbf{2.0183} & 2.2476 & 2.3311 & 2.2482
\\ & LPIPS & 0.0838 &  0.0718 & 0.0714 & \underline{0.0675} & 0.0685 & \textbf{0.0651} & 0.0677
\\
\hline
\end{tabular}
\end{center}
\end{table*}

\subsection{Training Details}
In training, we use the training set from DIV2K~\cite{timofte2017ntire} as the training set to train our models. The LR images are obtained by bicubic downsampling (BI) from the source high-resolution images. Images are augmented by rotating and flipping. We also randomly crop $48\times48$ patches from LR images as the input of the network.
Our networks are optimized with the ADAM optimizer~\cite{kingma2014adam}. The hyper-parameters $\beta_1$ and $\beta_2$ in the ADAM optimizer are set to $\beta_1 = 0.9$ and $\beta_2=0.999$. The batch size is set to 16.
The generator is pre-trained by the $L_1$ loss function, followed by generator and the discriminator training with the corresponding loss functions. 
Following~\cite{lim2017enhanced,lai2017deep,shi2016real,zhang2018residual,zhang2019rnan}, the initial learning rate is set to $1\times10^{-4}$, and then decays to half every $2\times10^5$ iterations. In FASRGAN, the coefficients in Eq.\ref{eq:generator_loss} are set as $\lambda_1 = 5e$-$3$, $\lambda_2 = 1e$-$2$ and $\lambda_3 = 1e$-$2$. Similar to ESRGAN~\cite{wang2018esrgan}, the number of RRDBs in the generator is set as 23. In Fs-SRGAN, we set the number of RRDBs in the shared feature extractor as $E=1$, and in the deep feature extractor as 16. The coefficients in Eq.\ref{eq:fs_generator} are set as $\lambda_1 = 5e$-$3$ and $\lambda_2 = 1e$-$2$. In FA+Fs-SRGAN, the number of RRDBs in the share part is set as 2, while in the deep feature extraction part is 15. The discriminator and the coefficients of the loss function are the same as those of FASRGAN. We implement our models with the PyTorch~\cite{paszke2017automatic} framework on two NVIDIA GeForce RTX 2080Ti GPUs.

\subsection{Datasets and Evaluation Metrics}
In the testing phase, we use seven standard benchmark datasets to evaluate the performance of our proposed methods:
Set5~\cite{bevilacqua2012low}, Set14~\cite{zeyde2010single}, BSD100~\cite{martin2001database}, Urban100~\cite{huang2015single}, Manga109~\cite{matsui2017sketch}, DIV2K validation~\cite{timofte2017ntire}, PIRM validation and test dataset\cite{Blau_2018_ECCV_Workshops}. Blau \emph{et al.}~\cite{blau2018perception} proved mathematically that perceptual quality is not always improved with the increase of PSNR value and there is a trade-off between average distortion and perceptual quality. Hence, we not only use PSNR and SSIM~\cite{wang2004image} to measure the reconstruction accuracy, but also adopt the learned perceptual image patch similarity (LPIPS)~\cite{zhang2018unreasonable} and  perceptual index (PI)~\cite{Blau_2018_ECCV_Workshops} to evaluate the perceptual quality of SR images. 
LPIPS firstly adopts a pre-trained network $\mathcal{F}$ to extract patches $y, y_0$ from the reference and target images $x, x_0$. The network $\mathcal{F}$ computes the activations of the image patches, each is scaled by a learned weight $w_l$ and then sums up the $L_2$ distances across all layers. Finally, it computes a perceptual real/fake prediction as follows:
\begin{equation}
    \label{eq:LPIPS}
    d(x, x_0) = \sum_l \frac{1}{H_lW_l}\sum_{h,w}\parallel w_l \odot (y_l^{hw} - \hat{y}^{hw}_{0l})\parallel_2^2,
\end{equation}
where $y_l, y_{0l} \in \mathbb{R}^{H_l \times W_l \times C_l}$ represent the reference or target features from layer $l$. The HR images from the public datasets are regarded as the reference images, the SR images generated by our methods or the compared methods as the target images. We use the public codes and pre-trained network (AlexNet from version 0.0) for evaluation. While PI is based on the non-reference
image quality measures of Ma \emph{et al.}~\cite{6353522} and NIQE~\cite{ma2017learning}: 
PI=$\frac{1}{2}$((10-Ma)+NIQE). PSNR and SSIM are calculated on the luminance channel in the YCbCr color space. We also use LPIPS and root mean square error (RMSE) to measure the trade-off between perceptual quality and reconstruction accuracy. Using LPIPS/RMSE rather than LPIPS/PSNR to evaluate the trade-off is for better observation, where lower LPIPS/RMSE means a better result. Higher PSNR/SSIM and lower RMSE mean better results in reconstruction accuracy, while lower scores of LPIPS/PI imply that the images are more similar to the HR images. 

As  shown in Fig.~\ref{fig:78004}, the SR image of our FASRGAN has less artifacts than that of ESRGAN~\cite{wang2018esrgan} and is clearer than that of RankSRGAN~\cite{zhang2019ranksrgan}. But the PI value of the SR image produced by RankSRGAN~\cite{zhang2019ranksrgan} is lower than that of our FASRGAN, and even lower than that of the original HR image. In terms of LPIPS, our method attains the lowest value, which is more consistent with human observation. Hence, we use LPIPS as our first perceptual quality metric and PI as the second one.


\begin{figure}
    \centering
    \includegraphics[width=1\linewidth]{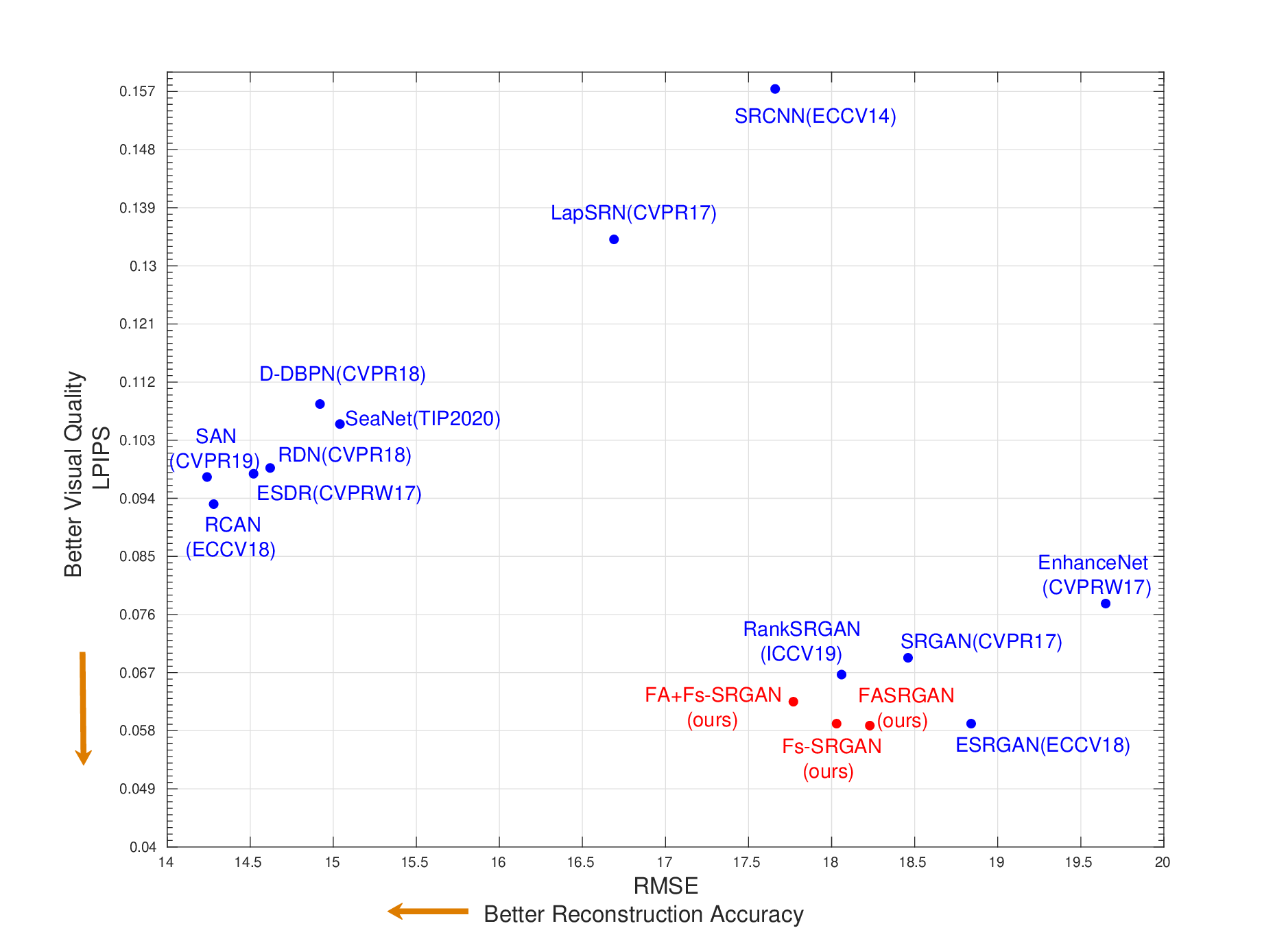}
    \caption{The trade-off of RMSE and LPIPS on Urban100 of our methods and the state-of-the-art methods for $4\times$ super-resolution.}
    \label{fig:PI_RMSE}
\end{figure}

\begin{figure*}[]
	\newlength\fsdurthree
	\setlength{\fsdurthree}{-0.4cm}
	\scriptsize
	\centering
	\begin{tabular}{cc}

		\begin{adjustbox}{valign=t}
		\tiny
			\begin{tabular}{c}
				\includegraphics[width=0.218\textwidth]{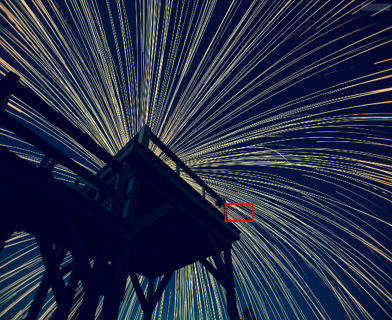} 
				\\
				DIV2K val ($4\times$): 
				\\
				0828
			\end{tabular}
		\end{adjustbox}
		\hspace{-0.4cm}
		\begin{adjustbox}{valign=t}
		\tiny
			\begin{tabular}{cccccc}
			    \centering
				\includegraphics[width=\widthscalefive \textwidth]{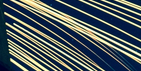} \hspace{\fsdurthree} &
				\includegraphics[width=\widthscalefive \textwidth]{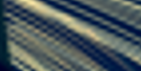}\hspace{\fsdurthree} &
				\includegraphics[width=\widthscalefive \textwidth]{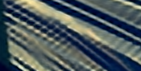} \hspace{\fsdurthree} &
				\includegraphics[width=\widthscalefive \textwidth]{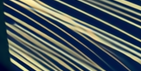} \hspace{\fsdurthree} &
				\includegraphics[width=\widthscalefive \textwidth]{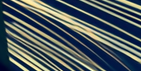} 
				\\
				HR \hspace{\fsdurthree} &
				Bicubic \hspace{\fsdurthree} &
				SRCNN~\cite{dong2015image} \hspace{\fsdurthree} &
				EDSR~\cite{lim2017enhanced} \hspace{\fsdurthree} &
				RCAN~\cite{zhang2018image} 
				\\
				PSNR/PI/LPIPS \hspace{\fsdurthree} &
				17.40/6.4557/0.2111 \hspace{\fsdurthree} &
				18.93/5.1946/0.1177 \hspace{\fsdurthree} &
				24.11/5.0693/0.0247 \hspace{\fsdurthree} &
				24.74/5.2368/0.0201
				\\
				\includegraphics[width=\widthscalefive \textwidth]{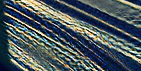} \hspace{\fsdurthree} &
				\includegraphics[width=\widthscalefive \textwidth]{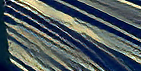} \hspace{\fsdurthree} &
				\includegraphics[width=\widthscalefive \textwidth]{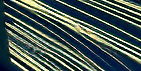} \hspace{\fsdurthree} &
				\includegraphics[width=\widthscalefive \textwidth]{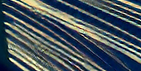} \hspace{\fsdurthree} &
				\includegraphics[width=\widthscalefive \textwidth]{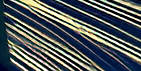} 
				\\
				SRGAN~\cite{ledig2017photo} \hspace{\fsdurthree} &
				EnhanceNet~\cite{sajjadi2017enhancenet} \hspace{\fsdurthree} &
				ESRGAN~\cite{wang2018esrgan} \hspace{\fsdurthree} &
				RankSRGAN~\cite{zhang2019ranksrgan} \hspace{\fsdurthree} &
				FASRGAN (ours) 
				\\
				21.24/3.9420/0.0407 \hspace{\fsdurthree} &
				19.57/4.0730/0.0562 \hspace{\fsdurthree} &
				22.09/3.8555/0.0230 \hspace{\fsdurthree} &
				21.35/3.6981/0.0366 \hspace{\fsdurthree} &
				22.39/3.9743/0.0214
				\\
			\end{tabular}
		\end{adjustbox}
		\vspace{0.5mm}
		
		\\
		
		\begin{adjustbox}{valign=t}
		\tiny
			\begin{tabular}{c}
				\includegraphics[width=0.218\textwidth]{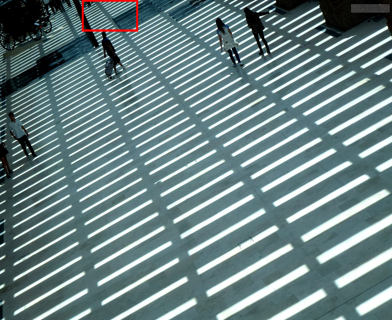} 
				\\
				Urban100 ($4\times$): 
				\\
				img\_093
			\end{tabular}
		\end{adjustbox}
		\hspace{-0.4cm}
		\begin{adjustbox}{valign=t}
		\tiny
			\begin{tabular}{cccccc}
			    \centering
				\includegraphics[width=\widthscalefive \textwidth]{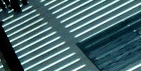} \hspace{\fsdurthree} &
				\includegraphics[width=\widthscalefive \textwidth]{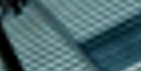}\hspace{\fsdurthree} &
				\includegraphics[width=\widthscalefive \textwidth]{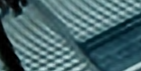} \hspace{\fsdurthree} &
				\includegraphics[width=\widthscalefive \textwidth]{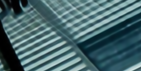}
				\hspace{\fsdurthree} &
				\includegraphics[width=\widthscalefive \textwidth]{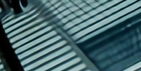}
				\\
				HR \hspace{\fsdurthree} &
				Bicubic \hspace{\fsdurthree} &
				SRCNN~\cite{dong2015image} \hspace{\fsdurthree} &
				EDSR~\cite{lim2017enhanced} \hspace{\fsdurthree} &
				RDN~\cite{zhang2018residual} 
				\\
				PSNR/PI/LPIPS \hspace{\fsdurthree} &
				23.62/7.2850/0.1711 \hspace{\fsdurthree} &
				26.17/5.4218/0.0594 \hspace{\fsdurthree} &
				29.54/6.0962/0.0262 \hspace{\fsdurthree} &
				28.59/5.9340/0.0252
				\\
				\includegraphics[width=\widthscalefive \textwidth]{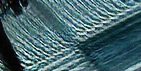} \hspace{\fsdurthree} &
				\includegraphics[width=\widthscalefive \textwidth]{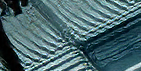} \hspace{\fsdurthree} &
				\includegraphics[width=\widthscalefive \textwidth]{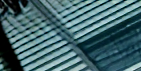} \hspace{\fsdurthree} &
				\includegraphics[width=\widthscalefive \textwidth]{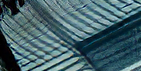} \hspace{\fsdurthree} &
				\includegraphics[width=\widthscalefive \textwidth]{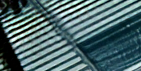}
				\\
				SRGAN~\cite{ledig2017photo} \hspace{\fsdurthree} &
				EnhanceNet~\cite{sajjadi2017enhancenet} \hspace{\fsdurthree} &
				ESRGAN~\cite{wang2018esrgan} \hspace{\fsdurthree} &
				RankSRGAN~\cite{zhang2019ranksrgan} \hspace{\fsdurthree} &
				FASRGAN (ours)
				\\
				26.56/4.4532/0.0236 \hspace{\fsdurthree} &
				24.96/4.2117/0.0464  \hspace{\fsdurthree} &
				28.49/4.9535/0.0181 \hspace{\fsdurthree} &
				25.93/4.1803/0.0284 \hspace{\fsdurthree} &
				28.17/4.8531/0.0188
				\\
			\end{tabular}
		\end{adjustbox}
		\vspace{0.5mm}
		
		\\

 	\end{tabular}
	\caption{
		The visual comparisons between FASRGAN and the state-of-the-art SR methods for $4\times$ super-resolution.
		}
\label{fig:visual_results_ASRGAN}
\end{figure*}

\subsection{Quantitative Comparisons}
We present the quantitative comparisons between our methods and the state-of-the-art perceptual image SR methods on several benchmark datasets. As shown in Table~\ref{tab:quantitative}, in most cases,  RankSRGAN~\cite{zhang2019ranksrgan} obtains the lowest PI values among these methods, benefiting from using the loss function with the newly added ranker to optimize the generator. However, our FASRGAN obtains the best LPIPS on most datasets, and both the LPIPS and PI values are better than that of ESRGAN~\cite{wang2018esrgan}, whose structure of the generator is the same as ours. It demonstrates that the fine-grained attention in our FASRGAN can transfer more information to the generator to produce better results. With less RRDBs in the generator, our Fs-SRGAN obtains best SSIM results and comparable, sometimes even better LPIPS results than those of ESRGAN~\cite{wang2018esrgan} and FASRGAN. In other words, our Fs-SRGAN extracts features more effectively and efficiently, benefiting from the feature-sharing mechanism. The combined model FA+Fs-SRGAN obtains the highest PSNR except for Set5, indicating that it can recover more contents in the SR images.

We also compare our methods with the state-of-the-art methods on the trade-off between reconstruction accuracy and visual quality. The results are shown in Fig.~\ref{fig:PI_RMSE}. 
Methods in the top-left part are almost MSE-based with low RMSE and high LPIPS scores due to the over-smoothed edges and lack of high-frequency details. The bottom-right category includes the GAN-based methods, such as SRGAN~\cite{ledig2017photo}, EnhanceNet~\cite{sajjadi2017enhancenet}, ESRGAN~\cite{wang2018esrgan}, RankSRGAN~\cite{zhang2019ranksrgan}, and our methods. These methods usually gain high-visual quality images even if their RMSE values are larger than those of the MSE-based methods. Our FASRGAN gets better visual quality and reconstruction accuracy compared with EnhanceNet, SRGAN and ESRGAN, and lower LPIPS than RankSRGAN. Our Fs-SRGAN attains comparable LPIPS with ESRGAN but lower RMSE, and better visual quality and reconstruction accuracy than RankSRGAN. The combined model FA+Fs-SRGAN obtains the lowest RMSE among the GAN-based methods. These demonstrate the effectiveness of our fine-grained attention and feature-sharing mechanism.

To further demonstrate the effectiveness of our FASRGAN and Fs-SRGAN, we conduct a user study to calculate the Mean Opinion Score (MOS)~\cite{lugmayr2019aim} against the state-of-the-art SR methods, i.e. SRGAN~\cite{ledig2017photo}, ESRGAN~\cite{wang2018esrgan} and RankSRGAN~\cite{zhang2019ranksrgan}. Ten candidates are shown with a side-by-side comparison of the generated SR image and the corresponding ground-truth. They are then asked to evaluate the difference of the two images on a 5-level scale defined as: 0 - `almost identical', 1 - `mostly similar’, 2 - `similar’, 3 - `somewhat similar’ and 4 - `mostly different’. 
We randomly select 10 images from PIRM val dataset~\cite{Blau_2018_ECCV_Workshops}, and invite 10 participants to give a score on each image according to the 5-level scale. For a better comparison, one small patch from the image is zoomed in. The average scores of all images are considered as the final results. As suggested in Table~\ref{tab:Mos}, our FASRGAN and Fs-SRGAN achieve better performance than all the compared methods, proving the effectiveness of our proposed fine-grained attention and feature-sharing mechanism.

\begin{table}[htbp]
\footnotesize
\centering
\begin{center}

\caption{The comparison of LPIPS and MOS between our methods and the state-of-the-art methods on PIRM Val, where the lower values mean more similar with the HR image. The LPIPS is tested on the whole dataset, while MOS is calculated on 10 randomly selected images.}
\label{tab:Mos}
\begin{tabular}{l||c|c}
\hline
\multirow{2}{*}{Methods} & \multicolumn{2}{c}{PIRM Val}
\\
\cline{2-3}
\multirow{2}{*}{}& LPIPS &  Mos  
\\ 
\hline
SRGAN~\cite{ledig2017photo} & 0.0718 & 1.98
\\ 
\hline
ESRGAN~\cite{wang2018esrgan} & 0.0714 & 1.88
\\ 
\hline
RankSRGAN~\cite{zhang2019ranksrgan} & \underline{0.0675} & \underline{1.84}
\\ 
\hline
 FASRGAN (ours) & 0.0685 & \textbf{1.46}
 \\ 
 \hline
 Fs-SRGAN (ours) & \textbf{0.0651} & \textbf{1.46}
\\
\hline
\end{tabular}
\end{center}
\end{table}


\begin{figure*}[]
	\setlength{\fsdurthree}{-0.4cm}
	\scriptsize
	\centering
	\begin{tabular}{cc}
		\begin{adjustbox}{valign=t}
		\tiny
			\begin{tabular}{c}
				\includegraphics[width=0.218\textwidth]{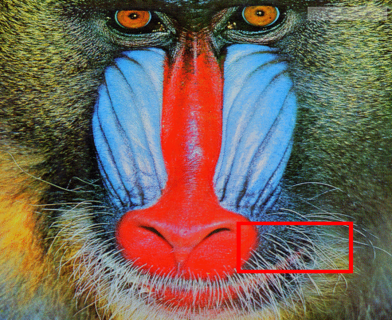}
				\\
				Set14 ($4\times$):
				\\
				baboon
			\end{tabular}
		\end{adjustbox}
		\hspace{-0.4cm}
		\begin{adjustbox}{valign=t}
		\tiny
			\begin{tabular}{cccccc}
			    \centering
				\includegraphics[width=\widthscalefive \textwidth]{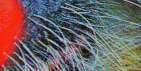} \hspace{\fsdurthree} &
				\includegraphics[width=\widthscalefive \textwidth]{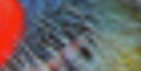}\hspace{\fsdurthree} &
				\includegraphics[width=\widthscalefive \textwidth]{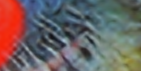} \hspace{\fsdurthree} &
				\includegraphics[width=\widthscalefive \textwidth]{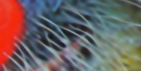} \hspace{\fsdurthree} &
				\includegraphics[width=\widthscalefive \textwidth]{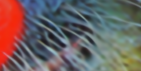}
				\\
				HR \hspace{\fsdurthree} &
				Bicubic \hspace{\fsdurthree} &
				SRCNN~\cite{dong2015image} \hspace{\fsdurthree} &
				EDSR~\cite{lim2017enhanced} \hspace{\fsdurthree} &
				RDN~\cite{zhang2018residual}
				\\
				PSNR/PI/LPIPS \hspace{\fsdurthree} &
				22.44/6.7571/0.3232 \hspace{\fsdurthree} &
				22.73/5.8182/0.2801 \hspace{\fsdurthree} &
				23.10/4.3197/0.2420 \hspace{\fsdurthree} &
				23.07/4.7250/0.2472
				\\
				\includegraphics[width=\widthscalefive
				\textwidth]{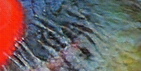} \hspace{\fsdurthree} &
				\includegraphics[width=\widthscalefive \textwidth]{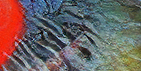} \hspace{\fsdurthree} &
				\includegraphics[width=\widthscalefive \textwidth]{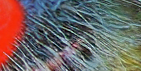} \hspace{\fsdurthree} &
				\includegraphics[width=\widthscalefive \textwidth]{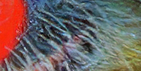} \hspace{\fsdurthree} &
				\includegraphics[width=\widthscalefive \textwidth]{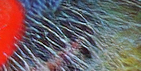}
				\\
				SRGAN~\cite{ledig2017photo} \hspace{\fsdurthree} &
				EnhanceNet~\cite{sajjadi2017enhancenet} \hspace{\fsdurthree} &
				ESRGAN~\cite{wang2018esrgan} \hspace{\fsdurthree} &
				RankSRGAN~\cite{zhang2019ranksrgan} \hspace{\fsdurthree} &
				Fs-SRGAN(ours)
				\\
				21.14/1.7589/0.1246 \hspace{\fsdurthree} &
				20.88/2.7198/0.1415 \hspace{\fsdurthree} &
				20.34/1.7318/0.0971 \hspace{\fsdurthree} &
				21.03/1.6523/0.1053 \hspace{\fsdurthree} &
				21.05/1.7846/0.1027
				\\
			\end{tabular}
		\end{adjustbox}
		\vspace{0.5mm}
		
		\\
		
		\begin{adjustbox}{valign=t}
		\tiny
			\begin{tabular}{c}
				\includegraphics[width=0.218\textwidth]{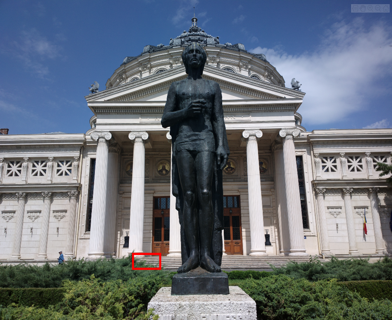}
				\\
				DIV2K val ($4\times$):
				\\
				0812
			\end{tabular}
		\end{adjustbox}
		\hspace{-0.4cm}
		\begin{adjustbox}{valign=t}
		\tiny
			\begin{tabular}{cccccc}
			    \centering
				\includegraphics[width=\widthscalefive \textwidth]{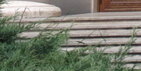} \hspace{\fsdurthree} &
				\includegraphics[width=\widthscalefive \textwidth]{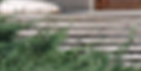}\hspace{\fsdurthree} &
				\includegraphics[width=\widthscalefive \textwidth]{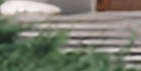} \hspace{\fsdurthree} &
				\includegraphics[width=\widthscalefive \textwidth]{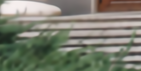}
				\hspace{\fsdurthree} &
		    	\includegraphics[width=\widthscalefive \textwidth]{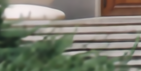}
				\\
				HR \hspace{\fsdurthree} &
				Bicubic \hspace{\fsdurthree} &
				SRCNN~\cite{dong2015image} \hspace{\fsdurthree} &
				EDSR~\cite{lim2017enhanced}
				\hspace{\fsdurthree} &
				RCAN~\cite{zhang2018image}
				\\
				PSNR/PI/LPIPS \hspace{\fsdurthree} &
				26.85/6.7023/0.2103 \hspace{\fsdurthree} &
				27.90/5.6650/0.1507 \hspace{\fsdurthree} &
				29.46/4.9867/0.1170 \hspace{\fsdurthree} &
				29.24/4.8927/0.1127
				\\
				\includegraphics[width=\widthscalefive \textwidth]{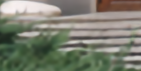} \hspace{\fsdurthree} &
				\includegraphics[width=\widthscalefive \textwidth]{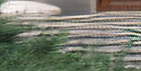} \hspace{\fsdurthree} &
				\includegraphics[width=\widthscalefive \textwidth]{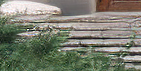} \hspace{\fsdurthree} &
				\includegraphics[width=\widthscalefive \textwidth]{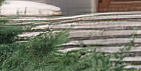} \hspace{\fsdurthree} &
				\includegraphics[width=\widthscalefive \textwidth]{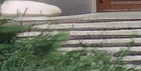}
				\\
				SRGAN~\cite{ledig2017photo} \hspace{\fsdurthree} &
				EnhanceNet~\cite{sajjadi2017enhancenet} \hspace{\fsdurthree} &
				ESRGAN~\cite{wang2018esrgan} \hspace{\fsdurthree} &
				RankSRGAN~\cite{zhang2019ranksrgan} \hspace{\fsdurthree} &
				Fs-SRGAN (ours)
				\\
				26.04/2.4359/0.0645 \hspace{\fsdurthree} &
				25.41/2.9509/0.0753 \hspace{\fsdurthree} &
				25.85/2.3963/0.0617 \hspace{\fsdurthree} &
				26.46/2.3470/0.0610 \hspace{\fsdurthree} &
				26.46/2.5413/0.0619
				\\
			\end{tabular}
		\end{adjustbox}
		\vspace{0.5mm}
		
		\\

 	\end{tabular}
	\caption{
		 The visual comparisons between Fs-SRGAN and the state-of-the-art SR methods for $4\times$.
	}
\label{fig:visual_results_Co-SRGAN}
\end{figure*}

\subsection{Qualitative Results}
We compare our final models on several public benchmark datasets with the state-of-the-art MSE-based methods: SRCNN~\cite{dong2015image}, EDSR~\cite{lim2017enhanced}, RDN~\cite{zhang2018residual},
RCAN~\cite{zhang2018image}, and GAN-based approaches: SRGAN~\cite{ledig2017photo}, EnhanceNet~\cite{sajjadi2017enhancenet}, ESRGAN~\cite{wang2018esrgan}, RankSRGAN~\cite{zhang2019ranksrgan}. 

\subsubsection{\textbf{Visual Comparisons of FASRGAN}}
Some representative quality results are presented in Fig.~\ref{fig:visual_results_ASRGAN}. PSNR , PI and LPIPS are also provided for reference.

As shown in Fig.~\ref{fig:visual_results_ASRGAN}, our proposed FASRGAN outperforms previous methods by a large margin. Images generated by FASRGAN contain more fine-grained textures and details. For example, 
the cropped parts of image `0828' 
are full of textures. All the compared MSE-based methods suffer from heavy blurry artifacts, failing to recover the structure and the gap of the stripes. SRGAN, EnhanceNet, ESRGAN and RankSRGAN generate high-frequency noise and wrong textures; while our FASRGAN can reduce noise and recover them more correctly, producing more faithful results and being closer to the HR images. For image `img\_093' in Urban100, the cropped parts of the images generated by the compared methods contain heavily blurry artifacts and lines with wrong directions. Although the LPIPS of ESRGAN is a little lower than our FASRGAN, our FASRGAN can alleviate the artifacts better and recover zebra crossing in the right direction. More results can be seen in the supplemental material. These comparisons demonstrate the strong ability of FASRGAN for producing more photo-realistic and high-quality SR images.

\subsubsection{\textbf{Visual Comparisons of Fs-SRGAN}}
We also compare our Fs-SRGAN with state-of-the-art methods in Fig.~\ref{fig:visual_results_Co-SRGAN}. Our Fs-SRGAN obtains better performance than other methods in producing SR images, in terms of sharpness and details. For image `baboon', the cropped parts of the images generated by the MSE-based methods are over-smoothed. Previous GAN-based methods not only fail to produce clear whiskers but also introduce lots of unpleasing noise. Despite having lower LPIPS value, ESRGAN generates too many whiskers, which have not appeared in the original HR image. While our Fs-SRGAN produces more correct whiskers. For image `0812'
,  MSE-based methods still suffer from heavy blurry artifacts and generate unnatural results. GAN-based methods cannot maintain the structures of the stairs or the train tracks and introduce artifacts. Our proposed Fs-SRGAN outperforms the compared methods, reducing the artifacts and recovering the correct textures. 
More results can be seen in the supplemental material.
These also indicate that the shared low-level feature extractor of the generator and the discriminator is beneficial.

\begin{figure*}[]
	\setlength{\fsdttwofigBD}{-0.4cm}
	\setlength{\fsdbigfig}{-0.38cm}
	\scriptsize
	\centering
	
	\begin{tabular}{c}	
		 \hspace{-2mm}
		\begin{adjustbox}{valign=t}
		\tiny
			\begin{tabular}{cccccc}
			   \includegraphics[width=0.138\textwidth]{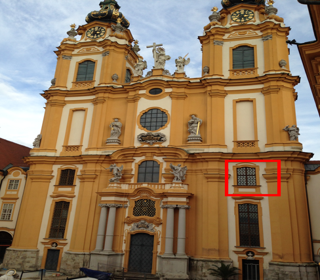} \hspace{\fsdbigfig}&
				\includegraphics[width=\widthscalefour \textwidth]{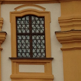} \hspace{\fsdttwofigBD} &
				\includegraphics[width=\widthscalefour \textwidth]{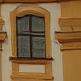} \hspace{\fsdttwofigBD} &
				\includegraphics[width=\widthscalefour \textwidth]{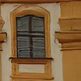} \hspace{\fsdttwofigBD} &
				\includegraphics[width=\widthscalefour \textwidth]{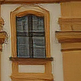} \hspace{\fsdttwofigBD} &
				\includegraphics[width=\widthscalefour \textwidth]{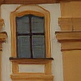} \hspace{\fsdttwofigBD}
				\\
    			PIRM val ($4\times$):  \hspace{\fsdbigfig} &
				HR \hspace{\fsdttwofigBD} &
    			ESRGAN ~\cite{wang2018esrgan}\hspace{\fsdttwofigBD} &
				FASRGAN (ours) \hspace{\fsdttwofigBD}
				&
				Fs-SRGAN (ours) \hspace{\fsdttwofigBD} &
				FA+Fs-SRGAN (ours) \hspace{\fsdttwofigBD}
				\\
				57 \hspace{\fsdbigfig} &
				PSNR/PI/LPIPS \hspace{\fsdbigfig} &
				25.03/2.3358/0.0563  \hspace{\fsdttwofigBD} &
				24.87/2.0577/0.0543  \hspace{\fsdttwofigBD} &
				25.19/2.0137/0.0586 \hspace{\fsdttwofigBD} &
				25.87/1.9691/0.0530  \hspace{\fsdttwofigBD}
				\\
			\end{tabular}
		\end{adjustbox}
        \vspace{0.5mm}
        \\
		\hspace{-2mm}
		\begin{adjustbox}{valign=t}
		\tiny
			\begin{tabular}{cccccc}
			   \includegraphics[width=0.138\textwidth]{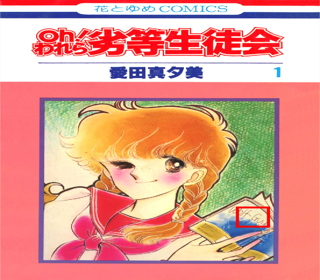} \hspace{\fsdbigfig}&
				\includegraphics[width=\widthscalefour \textwidth]{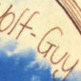} \hspace{\fsdttwofigBD} &
				\includegraphics[width=\widthscalefour \textwidth]{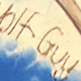} \hspace{\fsdttwofigBD} &
				\includegraphics[width=\widthscalefour \textwidth]{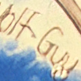} \hspace{\fsdttwofigBD} &
				\includegraphics[width=\widthscalefour \textwidth]{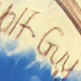} \hspace{\fsdttwofigBD} &
				\includegraphics[width=\widthscalefour \textwidth]{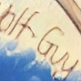} \hspace{\fsdttwofigBD}
				\\
    			Manga109 ($4\times$):  \hspace{\fsdbigfig} &
				HR \hspace{\fsdttwofigBD} &
    			ESRGAN ~\cite{wang2018esrgan}\hspace{\fsdttwofigBD} &
				FASRGAN (ours) \hspace{\fsdttwofigBD}
				&
				Fs-SRGAN (ours) \hspace{\fsdttwofigBD} &
				FA+Fs-SRGAN (ours) \hspace{\fsdttwofigBD}
				\\
				OhWareraRettouSeitokai \hspace{\fsdbigfig} &
				PSNR/PI/LPIPS \hspace{\fsdbigfig} &
				30.50/3.5038/0.0195    \hspace{\fsdttwofigBD} &
				29.62/3.4153/0.0227    \hspace{\fsdttwofigBD} &
				28.83/3.3394/0.0255    \hspace{\fsdttwofigBD} &
				29.75/3.4080/0.0188    \hspace{\fsdttwofigBD}
				\\
			\end{tabular}
		\end{adjustbox}
    \vspace{0.5mm}

 	\end{tabular}
	\caption{
		A visual results of FA+Fs-SRGAN for x4 magnification.
		}
	\label{fig:FA+Fs-SRGAN}
\end{figure*}

\subsubsection{\textbf{Visual Comparisons of FA+Fs-SRGAN}}
We further present the visual results of our FA+Fs-SRGAN compared with ESRGAN~\cite{wang2018esrgan}, FASRGAN and Fs-SRGAN. As shown in Fig.~\ref{fig:FA+Fs-SRGAN}, 
for image `57' and 'OhWareraRettouSeitokai', the results from FA+Fs-SRGAN are better than those of FASRGAN, and contain more correct textures. The PSNR and LPIPS values are both the best for FA+Fs-SRGAN. More results can be seen in the supplemental material. These results illustrate that the combined method can restore more contents for the SR images and obtains comparable or even better visual results compared with FASRGAN and Fs-SRGAN.

\subsection{Model Analysis}
This section compares the sizes and the time complexity of the generators between our methods and ESRGAN~\cite{wang2018esrgan}, which use RRDB as the basic block to construct the generators. We are not comparing our methods with SRGAN~\cite{ledig2017photo} and RankSRGAN~\cite{zhang2019ranksrgan} in these aspects as clearly they use 16 Resblocks to build their generators, so their parameters and inference time are less than ours.

In the aspect of the numbers of parameters, both ESRGAN and our FASRGAN have 23 RRDBs and 16.7M parameters, while our Fs-SRGAN and FA+Fs-SRGAN have 17 RRDBs and 12.46M parameters in their generators. 

In the aspect of inference time, we run our models and the public official test code and model from ESRGAN on Urban100 using a machine with 4.2GHz Inter i7 CPU (32G RAM) and Nvidia RTX 2080 platform. We conduct five times of inference on Urban100 and take the mean as the inference time. 

Our Fs-SRGAN and FA+Fs-SRGAN run much faster than ESRGAN, where Fs-SRGAN has the average time of 0.1377 seconds and FA+Fs-SRGAN 0.1364 seconds, while ESRGAN has 0.3573 seconds. Even our FASRGAN runs a little faster than ESRGAN, with average time 0.3160 seconds.
From Table~\ref{tab:quantitative} we can see that our Fs-SRGAN has comparable or even better results than ESRGAN, which demonstrates the efficiency of our feature-sharing mechanism.



\begin{figure}
  \centering
  \includegraphics[width=0.9\linewidth]{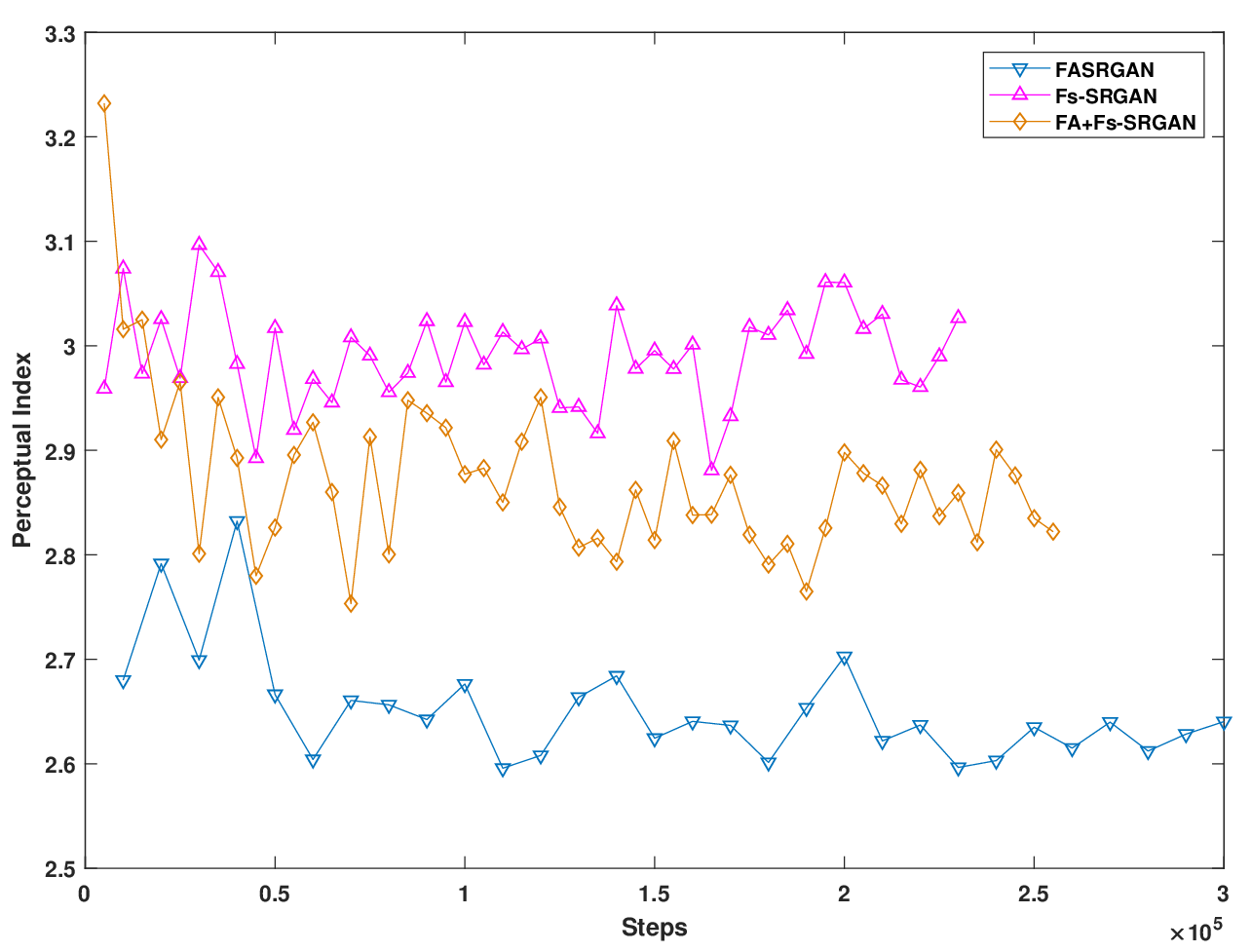}
  \caption{The changes of average PI on Set14 during the training process for $4\times$ super-resolution.}\label{fig:PI_process}
\end{figure}

Fig.~\ref{fig:PI_process} plots the curves of PI values in the training process of our proposed methods on Set14. We observe that the training process of FASRGAN is more stable and the PI value is the lowest. 
The average PI value of Fs-SRGAN is higher than FASRGAN. 
As described in~\cite{wang2018esrgan}, the deep model has a stronger representation capacity to capture semantic information and reduce unpleasing noises. And as mentioned above, Fs-SRGAN contains fewer RRDBs than FASRGAN. 
Hence, we speculate that compared with FASRGAN, Fs-SRGAN with fewer RRDBs captures less information for reconstruction but brings more noises, causing higher PI values. FA$+$Fs-SRGAN, which combines the fine-grained attention mechanism into Fs-SRGAN, obtains the lower PI values than Fs-SRGAN, which  demonstrates the effectiveness of our fine-grained attention mechanism. 
However, the training of the FA+Fs-SRGAN is not stable, which is the concern we need to focus on in our future work.

\begin{table}[htbp]
\footnotesize
\centering
\begin{center}

\caption{The result of object recognition between our methods and the state-of-the-art methods for $4\times$ SR. The baseline uses the original HR image as the input of ResNet-50 model.}
\label{tab:object_recognition_SRGAN}

\begin{tabular}{l|c|c}
\hline
Evaluation  & Top-1 error& Top-5 error
\\ \hline
Bicubic  & 0.526 & 0.277
\\ \hline
SRCNN~\cite{dong2015image} & 0.464 & 0.230
\\ \hline
FSRCNN~\cite{dong2016accelerating} & 0.488 & 0.252
\\ \hline
SRGAN~\cite{ledig2017photo} & 0.410 & 0.191
\\ \hline
EnhanceNet~\cite{sajjadi2017enhancenet} & 0.454 & 0.224
\\ \hline
ESRGAN~\cite{wang2018esrgan} & \underline{0.334}  & \underline{0.132}
\\ \hline
RankSRGAN~\cite{zhang2019ranksrgan} & 0.342 & 0.136
\\ \hline
Fs-SRGAN (ours) &0.338 &0.136
\\ \hline
FA+Fs-SRGAN (ours)  &0.337  &0.134
\\ \hline
FASRGAN (ours)  & \textbf{0.323} & \textbf{0.124}
\\ \hline
Baseline & 0.241 & 0.071
\\
\hline
\end{tabular}
\end{center}
\end{table}

\subsection{Object Recognition Performance}

To further demonstrate the quality of our generated SR images, we treat them as a pre-processing step for object recognition.
We use the same setting as EnhanceNet and evaluate the object recognition performance with the generated images by our methods and other state-of-the-art methods: SRCNN~\cite{dong2015image}, FSRCNN~\cite{dong2016accelerating}, SRGAN~\cite{ledig2017photo}, EnhanceNet~\cite{sajjadi2017enhancenet}, ESRGAN~\cite{wang2018esrgan}, RankSRGAN~\cite{zhang2019ranksrgan}.

We use the pre-trained ResNet-50 on imageNet as an evaluation model and fetch the first 1000 images in ImageNet CLS-LOC validation dataset for evaluation. The test images are first down-sampled by bicubic and then upscaled by our methods and the compared methods. These SR images are then used as inputs to the ResNet-50 model to calculate their Top-1 and Top-5 errors for evaluation. As shown in Table~\ref{tab:object_recognition_SRGAN}, 
both two methods we proposed and the variant FA$+$Fs-SRGAN achieve considerable accuracy compared to the state-of-the-art methods. Among these three methods, FASRGAN achieves the lowest Top-1 and Top-5 errors, Fs-SRGAN and FA+Fs-SRGAN obtain comparable results with ESRGAN, demonstrating the effectiveness of both the fine-grained attention and the feature-sharing mechanisms.



\begin{table}[htbp]
\footnotesize
\centering
\begin{center}

\caption{The ablation study of fine-grained attention (FA) and feature-sharing (Fs) mechanisms for $4\times$ super-resolution.}
\label{tab:ablation_result}

\begin{tabular}{c|c|c|c||c|c}

\hline
\multicolumn{2}{c|}{\multirow{2}{*}{Model}} & \multicolumn{2}{c||}{FA mechanism} & \multicolumn{2}{c}{Fs mechanism}
\\
\cline{3-6}
\multicolumn{2}{c|}{}& w/o FA & FASRGAN & w/o Fs & Fs-SRGAN
\\
 \hline
 & PSNR & 25.04 & 25.26 & 25.44 & 25.69
 \\
 PIRM & SSIM & 0.6454 & 0.6523 & 0.6626 & 0.6785
 \\
 Test & PI & 2.4251  & 2.1160 &  2.1420 & 2.2279
 \\
 & LPIPS & 0.0751 & \textbf{0.0718} & 0.0731 & \textbf{0.0695}
 \\ 
 \hline
\end{tabular}
\end{center}
\end{table}

\begin{figure*}[]
	\setlength{\fsdttwofigBD}{-0.4cm}
	\setlength{\fsdbigfig}{-0.38cm}
	\scriptsize
	\centering
	\begin{tabular}{c}
		\hspace{-2mm}
		\begin{adjustbox}{valign=t}
		\tiny
			\begin{tabular}{ccccc}
			    \includegraphics[width=0.148\textwidth]{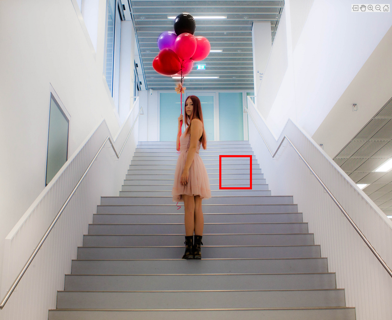} \hspace{\fsdbigfig}&
				\includegraphics[width=\widthscalefour \textwidth]{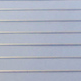} \hspace{\fsdttwofigBD} &
				\includegraphics[width=\widthscalefour \textwidth]{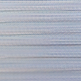} \hspace{\fsdttwofigBD} &
				\includegraphics[width=\widthscalefour \textwidth]{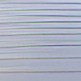} \hspace{\fsdttwofigBD} &
				\includegraphics[width=\widthscalefour \textwidth]{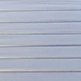}
				\\
				Urban100(x4):\hspace{\fsdbigfig} &
				HR \hspace{\fsdttwofigBD} &
				SRGAN~\cite{ledig2017photo}\hspace{\fsdttwofigBD} &
				w/o Attention \hspace{\fsdttwofigBD} &
				FASRGAN (ours) 
				\\
				img\_009 \hspace{\fsdbigfig} &
				PSNR/PI/LPIPS \hspace{\fsdttwofigBD} &
				32.51/3.6891/0.0360 \hspace{\fsdttwofigBD} &
				31.11/3.6991/0.0384 \hspace{\fsdttwofigBD} &
				32.13/3.6700/0.0345

				\\
			\end{tabular}
		\end{adjustbox}
        \vspace{0.5mm}
	\\
	
		\hspace{-2mm}
		\begin{adjustbox}{valign=t}
		\tiny
			\begin{tabular}{ccccc}
			    \includegraphics[width=0.148\textwidth]{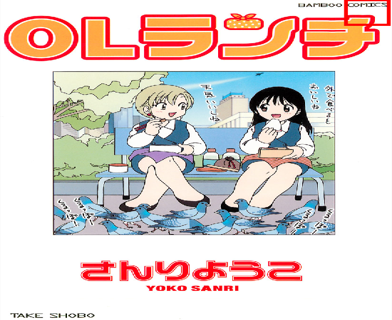} \hspace{\fsdbigfig}&
				\includegraphics[width=\widthscalefour \textwidth]{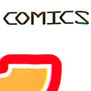} \hspace{\fsdttwofigBD} &
				\includegraphics[width=\widthscalefour \textwidth]{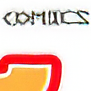} \hspace{\fsdttwofigBD} &
				\includegraphics[width=\widthscalefour \textwidth]{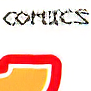} \hspace{\fsdttwofigBD} &
				\includegraphics[width=\widthscalefour \textwidth]{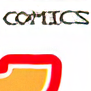}
				\\
				Manga109(x4):\hspace{\fsdbigfig} &
				HR \hspace{\fsdttwofigBD} &
				SRGAN~\cite{ledig2017photo}\hspace{\fsdttwofigBD} &
				w/o Attention \hspace{\fsdttwofigBD} &
				FASRGAN (ours) 
				\\
				OL\_Lunch \hspace{\fsdbigfig} &
				PSNR/PI/LPIPS \hspace{\fsdttwofigBD} &
				24.02/3.5644/0.0291 \hspace{\fsdttwofigBD} &
				24.19/3.9100/0.0207 \hspace{\fsdttwofigBD} &
				24.18/3.7275/0.0222 
			
				\\
				
			\end{tabular}
		\end{adjustbox}
    \vspace{0.5mm}

	\end{tabular}
	\caption{
		The visual results of ablation study of FASRGAN $4\times$ SR.
	}
\label{fig:ablation_FASRGAN}
\end{figure*}

\begin{figure*}[]
	\setlength{\fsdttwofigBD}{-0.4cm}
	\setlength{\fsdbigfig}{-0.38cm}
	\scriptsize
	\centering
	\begin{tabular}{c}

		\hspace{-2mm}
		\begin{adjustbox}{valign=t}
		\tiny
			\begin{tabular}{ccccc}
			    \includegraphics[width=0.148\textwidth]{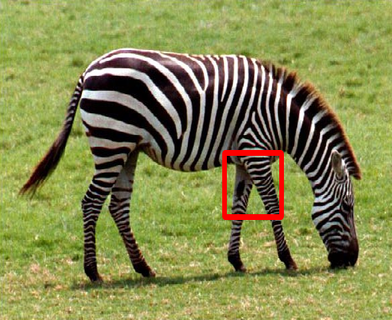} \hspace{\fsdbigfig}&
				\includegraphics[width=\widthscalefour
			     \textwidth]{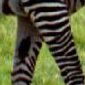} \hspace{\fsdttwofigBD} &
				\includegraphics[width=\widthscalefour \textwidth]{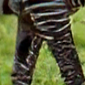} \hspace{\fsdttwofigBD} &
				\includegraphics[width=\widthscalefour \textwidth]{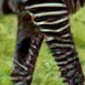} \hspace{\fsdttwofigBD} &
				\includegraphics[width=\widthscalefour \textwidth]{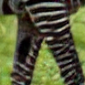}
				\\
				Set14(x4): \hspace{\fsdbigfig} &
				HR \hspace{\fsdttwofigBD} &
				SRGAN~\cite{ledig2017photo} \hspace{\fsdttwofigBD} &
				w/o Feature-sharing \hspace{\fsdttwofigBD} &
				Fs-SRGAN (ours) \hspace{\fsdttwofigBD}
				\\
				zebra \hspace{\fsdbigfig} &
				PSNR/PI/LPIPS \hspace{\fsdttwofigBD} &
				24.81/3.1552/0.0681 \hspace{\fsdttwofigBD} &
				25.00/3.0571/0.0572 \hspace{\fsdttwofigBD} &
				25.71/3.1393/0.0582 \hspace{\fsdttwofigBD}
				\\
			\end{tabular}
		\end{adjustbox}
    \vspace{0.5mm}
	
	\\
	
		\hspace{-2mm}
		\begin{adjustbox}{valign=t}
		\tiny
			\begin{tabular}{ccccc}
			    \includegraphics[width=0.148\textwidth]{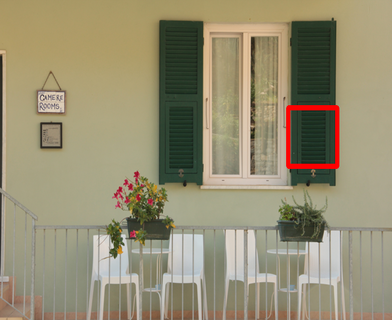} \hspace{\fsdbigfig}&
				\includegraphics[width=\widthscalefour \textwidth]{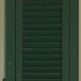} \hspace{\fsdttwofigBD} &
				\includegraphics[width=\widthscalefour \textwidth]{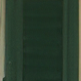} \hspace{\fsdttwofigBD} &
				\includegraphics[width=\widthscalefour \textwidth]{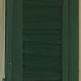} \hspace{\fsdttwofigBD} &
				\includegraphics[width=\widthscalefour \textwidth]{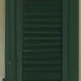}
				\\
                PIRM Test(x4): \hspace{\fsdbigfig} &
				HR \hspace{\fsdttwofigBD} &
				SRGAN~\cite{ledig2017photo} \hspace{\fsdttwofigBD} &
				w/o feature-sharing \hspace{\fsdttwofigBD} &
				Fs-SRGAN (ours) \hspace{\fsdttwofigBD}
				\\
				292 \hspace{\fsdbigfig} &
				PSNR/PI/LPIPS \hspace{\fsdttwofigBD} &
				31.06/3.8133/0.0432 \hspace{\fsdttwofigBD} &
				30.34/3.8192/0.0498 \hspace{\fsdttwofigBD} &
				30.84/3.9621/0.0398 \hspace{\fsdttwofigBD}
				\\
			\end{tabular}
		\end{adjustbox}
    \vspace{0.5mm}
    \\

	\end{tabular}
	\caption{
		The visual results of ablation study of Fs-SRGAN for $4\times$ SR.
	}
\label{fig:ablation_Co-SRGAN}
\end{figure*}


\subsection{Ablation Study}
To study the effects of the two mechanisms in the proposed methods, we conduct ablation experiments by removing the mechanisms and test the differences, respectively. The quantitative results are illustrated in Table~\ref{tab:ablation_result}, overall visual comparisons are presented in Fig.~\ref{fig:ablation_FASRGAN}, Fig.~\ref{fig:ablation_Co-SRGAN} and the supplemental material. A detailed discussion is provided as follows.

\subsubsection{\textbf{Removing the Fine-grained Attention Mechanism}}
We first remove the fine-grained attention (FA) mechanism in FASRGAN. The attention item is removed from the loss functions of the generator in the model without FA. The coefficients of Eq.\ref{eq:generator_loss} are set as $\lambda_1 = 5e$-$3$, $\lambda_2 = 0$ and $\lambda_3 = 1e$-$2$. The fine-grained adversarial loss functions, $L^D_M$ and $L^G_{fine}$ are also removed. The generators of FASRGAN and the model without FA have the same parameters, and the difference lies in the loss function in training. 

From Table~\ref{tab:ablation_result} we can see that FASRGAN surpasses the model without FA in all metrics. An obvious performance decrease can be observed in Fig.~\ref{fig:ablation_FASRGAN}. For image ‘img\_009’,
the model without FA mechanism introduces some unnatural noises and undesired edges, while FASRGAN can maintain the structure and produce high-quality SR images. 
For image ‘OL\_Lunch’, the result from the model without FA mechanism contains more artifacts and noise and the letters cannot be well recognized, while FASRGAN reduces the artifacts and noises, whose result looks closer to the original HR images. 
The visual analysis indicates the effectiveness of the FA mechanism in removing unpleasant and unnatural artifacts.

\subsubsection{\textbf{Removing the Feature-sharing Mechanism}}
We remove the feature-sharing (Fs) mechanism, so that the generator and discriminator extract their low-level features separately, but the loss function keeps the same as that of Fs-SRGAN. The discriminator and the generator in our Fs-SRGAN use a shared RRDB to extract low-level features, while in the case the Fs mechanism is removed, different RRDBs are used to extract low-level features for them individually.  

Table~\ref{tab:ablation_result} shows that Fs-SRGAN has lower LPIPS and higher PSNR/SSIM than the model without Fs mechanism. Fig.~\ref{fig:ablation_Co-SRGAN} presents the results of the model without Fs mechanism and Fs-SRGAN. We can observe that Fs-SRGAN outperforms the model without Fs mechanism by a large margin. The removal of Fs mechanism tends to introduce unpleasant artifacts. For image `zebra', by employing the Fs mechanism, Fs-SRGAN can alleviate heavy artifacts and noises, recovering the strips of legs more clearly and correctly. 
For image `292', our Fs-SRGAN generates more textures of the pane. The above results illustrate the effectiveness of our Fs mechanism.

\subsubsection{\textbf{Feature Visualization of Fine-Grained Attention and Feature-sharing Mechanisms}}
\begin{figure*}[]
	\setlength{\fsdttwofigBD}{-0.4cm}
	\setlength{\fsdbigfig}{-0.38cm}
	\scriptsize
	\centering
	\begin{tabular}{c}
		\hspace{-2mm}
		\begin{adjustbox}{valign=t}
		\tiny
			\begin{tabular}{ccc|ccc}
			    \includegraphics[width=\widthscalefour \textwidth]{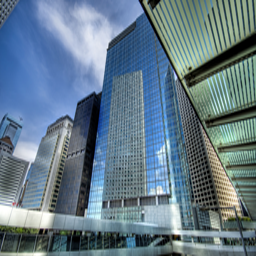} \hspace{\fsdbigfig}&
				\includegraphics[width=\widthscalefour \textwidth]{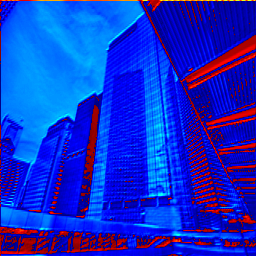} \hspace{\fsdttwofigBD} &
				\includegraphics[width=\widthscalefour\textwidth]{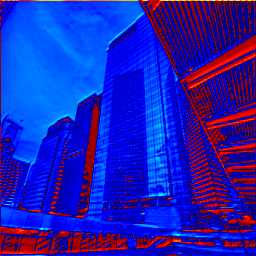} \hspace{\fsdbigfig} \hspace{3mm} &
		        \includegraphics[width=\widthscalefour \textwidth]{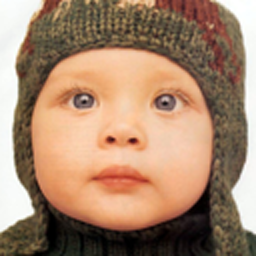} \hspace{\fsdbigfig}&
				\includegraphics[width=\widthscalefour\textwidth]{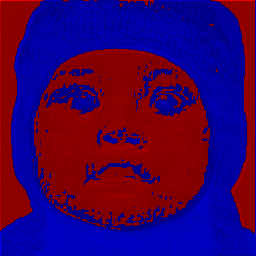} \hspace{\fsdttwofigBD} &
				\includegraphics[width=\widthscalefour\textwidth]{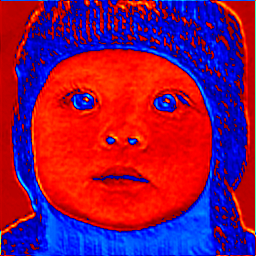}
				\\
				Urban100: \hspace{\fsdbigfig} &	
				w/o Attention \hspace{\fsdttwofigBD} &
				FASRGAN (ours)  \hspace{\fsdbigfig} \hspace{3mm} &
				Set5: \hspace{\fsdbigfig} &	
				w/o Feature-sharing
				\hspace{\fsdttwofigBD} &
				Fs-SRGAN (Ours)
				\\
				img\_061 \hspace{\fsdbigfig} &
				the 7-th channel \hspace{\fsdttwofigBD} &
				the 7-th channel \hspace{\fsdbigfig} \hspace{3mm} &
				baby \hspace{\fsdbigfig} &
				the 11-th channel \hspace{\fsdttwofigBD} &
				the 11-th channel
			\end{tabular}
		\end{adjustbox}
        \vspace{0.5mm}
	\\
	\hspace{-2mm}
		\begin{adjustbox}{valign=t}
		\tiny
			\begin{tabular}{ccc|ccc}
			    \includegraphics[width=\widthscalefour \textwidth]{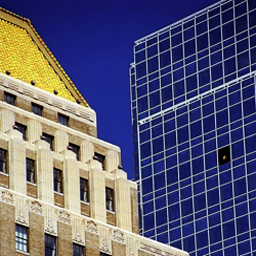} \hspace{\fsdbigfig}&
				\includegraphics[width=\widthscalefour\textwidth]{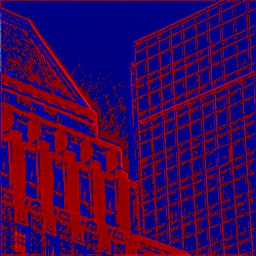} \hspace{\fsdttwofigBD} &
				\includegraphics[width=\widthscalefour\textwidth]{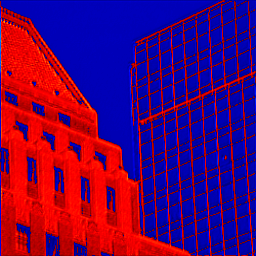} \hspace{\fsdbigfig} \hspace{3mm} &
			    \includegraphics[width=\widthscalefour\textwidth]{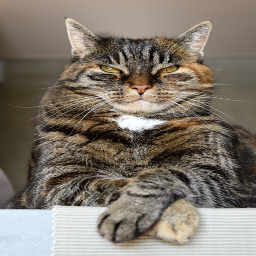} \hspace{\fsdbigfig}&
		    	\includegraphics[width=\widthscalefour\textwidth]{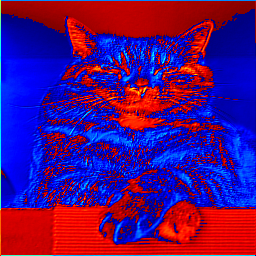} \hspace{\fsdttwofigBD} &
				\includegraphics[width=\widthscalefour\textwidth]{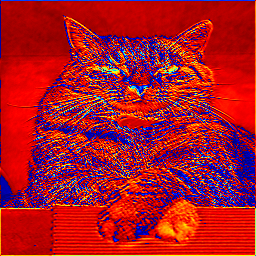}
				\\
				Urban100: \hspace{\fsdbigfig} &
				w/o Attention \hspace{\fsdttwofigBD} &
				FASRGAN (ours)  \hspace{\fsdttwofigBD} \hspace{3mm} &
				DIV2K val: \hspace{\fsdbigfig} &
				w/o Feature-sharing
				\hspace{\fsdttwofigBD} &
				Fs-SRGAN (Ours)
				\\
				img\_063 \hspace{\fsdbigfig} &
				the 15-th channel \hspace{\fsdttwofigBD} &
				the 15-th channel \hspace{\fsdbigfig} \hspace{3mm} &
		        0869 \hspace{\fsdbigfig} &
				the 6-th channel \hspace{\fsdttwofigBD} &
				the 6-th channel
			\end{tabular}
		\end{adjustbox}
        \vspace{0.5mm}
	\end{tabular}
	\caption{
		The feature visualization of the first RRDB of the generators in FASRGAN and the model w/o attention, and of the shared low-level feature extractor in Fs-SRGAN and the first RRDBs of the generator in the model w/o feature-sharing. 
	}
\label{fig:fea_visualization}
\end{figure*}

To further verify the effectiveness of our proposed fine-grained attention and feature-sharing mechanisms, we present feature visualizations of the first RRDB of the generator (FASRGAN) and the shared feature extraction part (Fs-SRGAN) in Fig.\ref{fig:fea_visualization}. FASRGAN reduces the noises in the image `img\_063' and extracts more texture information compared with the model without attention mechanism. The feature maps from Fs-SRGAN also contain more helpful textures, showing that our proposed methods help the networks extract more useful information.

\subsubsection{\textbf{The Block Number of the Feature-sharing Part}}
\begin{figure}
  \centering

  \includegraphics[width=0.9\linewidth]{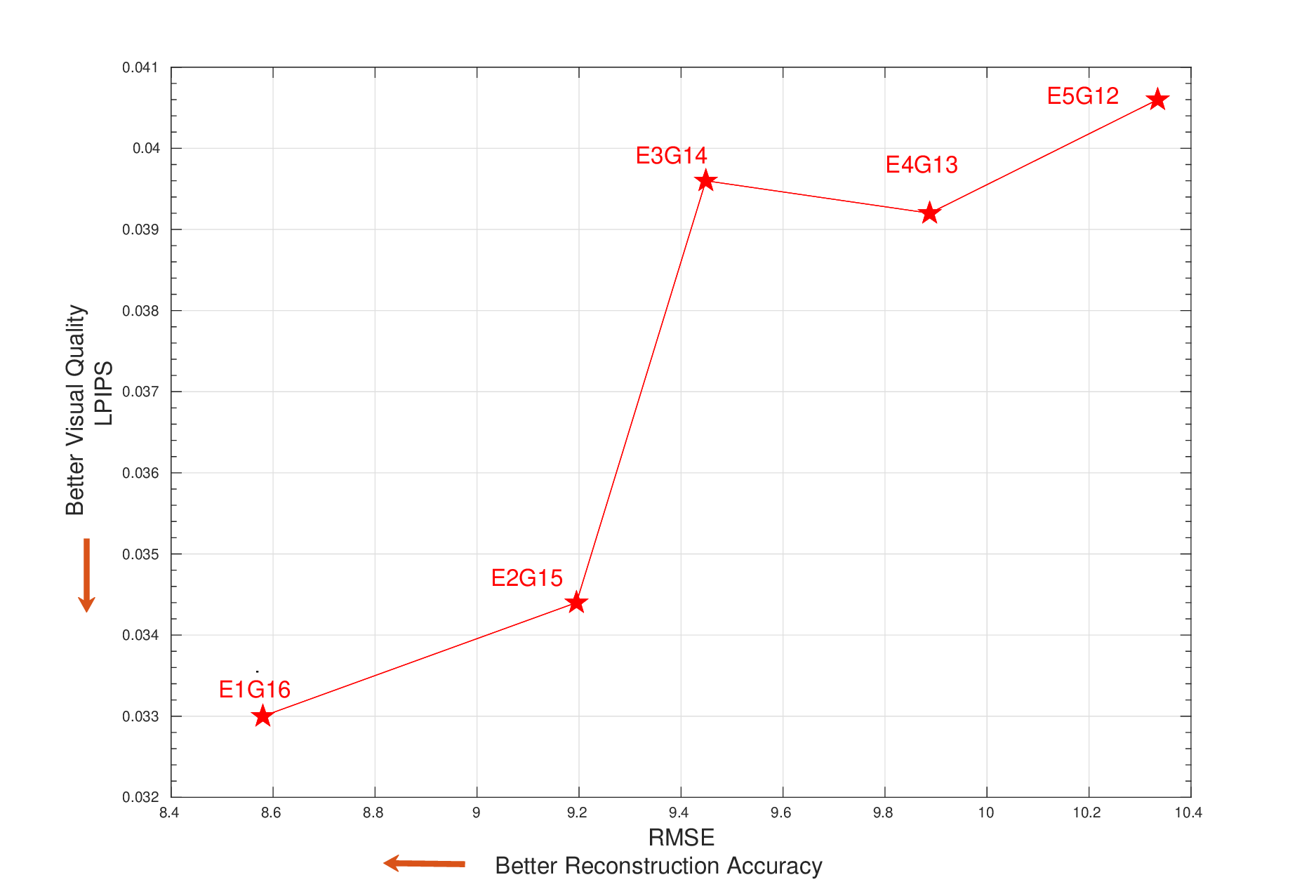}
  \caption{The change of the number of RRDBs in shared low-level feature extractor (E) in Fs-SRGAN. G represents the number of RRDBs in the deep feature extraction part. The test is conducted on Set5 for $4\times$ super-resolution.}\label{fig:change}
\end{figure}

To further study the effect of the depth of the shared feature extractor in Fs-SRGAN, we vary the number of RRDBs in both the shared low-level feature extractor and the deep feature extractor, keeping the total number of RRDBs in the generator unchanged. As shown in Fig.~\ref{fig:change}, increasing the number of the shared feature extractor leads to performance reduction and increases the burden of the discriminator due to more parameters which makes the model difficult to train. Among them, E1G16 obtains the best results in both visual quality and reconstruction accuracy.

\begin{table}[htbp]
\footnotesize
\centering
\begin{center}

\caption{The ablation study of the coefficient $\lambda_2$ of $L_{attention}$ in Eq.\ref{eq:atten}}
\label{tab:coefficient}

\begin{tabular}{c|c|c|c|c}

\hline
Dataset & Metric & $\lambda_2 = 0.05$ & $\lambda_2 =0.01$ & $\lambda_2 =0.005$ 
\\ \hline
\multirow{4}{*}{Urban100} & PSNR & \underline{24.35} & \textbf{24.51} & 24.31
\\
& SSIM &0.7359  & \textbf{0.7380} & \underline{0.7364}
\\
& PI & \textbf{3.5091} & \underline{3.5173} & 3.5284
\\
& LPIPS & \underline{0.0608} & \textbf{0.0588} & 0.0613
\\
\hline
\multirow{4}{*}{DIV2K val} & PSNR & \underline{28.16} & 28.15 & \textbf{28.19}
\\
& SSIM & 0.7771 & \textbf{0.7903} & \underline{0.7803}
\\
& PI & \textbf{3.1826} & 3.3303 & \underline{3.2378}
\\
& LPIPS & \underline{0.0547} & \textbf{0.0542} & 0.0560
\\
\hline

\end{tabular}
\end{center}
\end{table}

\subsubsection{\textbf{Coefficient of the Fine-Grained Attention Loss in the FASRGAN Generator}}
We also conduct an ablation study to verify the influences of different coefficient $\lambda_2$ of the fine-grained attention loss $L_{attention}$ in the generator of FASRGAN. We set $\lambda_2$ as 0.05, 0.01 and 0.005, while the other settings are kept the same. As shown in Table~\ref{tab:coefficient}, the model with $\lambda_2 = 0.01$ has the best performance in SSIM and LPIPS on Urban100 and DIV2K val, and achieves comparable results in PSNR and PI. The visual results are shown in the supplemental material. When $\lambda_2$ is set too small, the fine-grained feedback from the discriminator has less impact on the generator. And when $\lambda_2$ is set too large, the training is unstable for both the generator and discriminator and hard to converge. These results indicate that $\lambda_2 = 0.01$ is a good setting in practice, which is used in our FASRGAN.

\begin{table}[htbp]
\footnotesize
\centering
\begin{center}

\caption{The effect of fine-grained attention (FA) and feature-sharing (Fs) mechanisms on SRGAN.}
\label{tab:srgan}

\begin{tabular}{c|c|c|c|c}

\hline
Dataset & Metric & SRGAN~\cite{ledig2017photo} & SRGAN\_FA & SRGAN\_Fs 
\\ \hline
\multirow{4}{*}{Set5} & PSNR & \textbf{29.91}  & 29.61   & \underline{29.66}
\\
& SSIM & \underline{0.8510}  & 0.8437  & \textbf{0.8541}
\\
& PI & \underline{3.4322} &  \textbf{3.0651} & 3.4440
\\
& LPIPS &0.0389  & \textbf{0.0341} & \underline{0.0368}
\\ \hline
\multirow{4}{*}{Set14} & PSNR & \textbf{26.56}  &  26.11 & \underline{26.27}
\\
& SSIM & \underline{0.7093} & 0.6977 & \textbf{0.7179}
\\
& PI & 	2.8549  & \textbf{2.7550} & \underline{2.7705}
\\
& LPIPS &0.0696  & \underline{0.0692} & \textbf{0.0669}
\\ \hline
\multirow{4}{*}{Urban100} & PSNR & \textbf{24.39}  & 24.00 & \underline{24.04}
\\
& SSIM & \underline{0.7309} & 0.7205 & \textbf{0.7331}
\\
& PI & \underline{3.4814} & \textbf{3.4252}  & 3.4818
\\
& LPIPS &0.0693  & \textbf{0.0688} & \underline{0.0691}
\\
\hline
\multirow{4}{*}{BSD100} & PSNR &  \textbf{25.50}   & 25.35 & \underline{25.46}
\\
& SSIM & \underline{0.6528} &  0.6506 & \textbf{0.6650}
\\
& PI & 	\underline{2.3054}  & \textbf{2.2503} & 2.3348
\\
& LPIPS & 	0.0887 & \textbf{0.0856} & \underline{0.0876}
\\
\hline

\end{tabular}
\end{center}
\end{table}

\subsubsection{\textbf{The Fine-grained Attention and Feature-sharing Mechanisms in SRGAN}}
To verify whether our proposed fine-grained attention (FA) and feature-sharing mechanisms can improve the performance in other GAN-based SR models, we incorporate these two mechanisms into SRGAN~\cite{ledig2017photo}, denoting as SRGAN\_FA and SRGAN\_Fs respectively. The generator in SRGAN\_FA is the same as that of SRGAN~\cite{ledig2017photo}, and the discriminator adopts our proposed UNet-like structure. We use a convolution layer and a residual block (RB) as the shared low-level feature extraction part for the generator and discriminator in SRGAN\_Fs. The rest part of the generator and the discriminator are similar with that of SRGAN~\cite{ledig2017photo}, except that the number of RB in the deep feature extraction is 13. Hence, the parameter for SRGAN\_Fs is 1.48K, and 1.554K for SRGAN and SRGAN\_FA. As shown in Table~\ref{tab:srgan}, SRGAN\_FA achieves the best performance in terms of PI and LPIPS on most of the test dataset. SRGAN\_Fs also outperforms SRGAN in SSIM and LPIPS. These results demonstrate that our proposed FA and Fs mechanisms can be well adapted to the SRGAN model.

\begin{table}[htbp]
\scriptsize
\footnotesize

\begin{center}

\caption{The quantitative comparison of our methods and other SR methods on RealSR for $4\times$ magnification. FT represents the model fine-tunes on RealSR training dataset.}
\label{tab:realsr}

\begin{tabular}{c|c|c|c|c}

\hline
\multirow{2}{*}{Model} & \multicolumn{4}{c}{RealSR (V3) Test}
\\
\cline{2-5}
 & PSNR & SSIM & PI & LPIPS
\\
 \hline
SRCNN~\cite{dong2015image} & \textbf{27.69} & \textbf{0.7808}  & 7.8679 & 0.2290
\\
RCAN ~\cite{zhang2018image} & \underline{27.65} & \underline{0.7803} & 7.8519 & 0.2311 
\\
ESRGAN~\cite{wang2018esrgan} & 27.57 & 0.7748 & 7.4819 & 0.2215
\\
RankSRGAN~\cite{zhang2019ranksrgan} & 27.56 & 0.7701 & \textbf{7.0521} & \textbf{0.2100}
\\
FASRGAN (ours) & 27.57  & 0.7732 & \underline{7.1178} & \underline{0.2111}
\\
Fs-SRGAN (ours) & 27.47 & 0.7744 & 7.3112 & 0.2151 

\\
\hline
\hline
ZSSR~\cite{shocher2018zero} & \underline{27.56} & 0.7719 & 7.4666 & 0.2069
\\
FSSR~\cite{fritsche2019frequency} & 26.68 & \underline{0.7773} & 7.0811 & 0.1978
\\
ESRGAN~\cite{wang2018esrgan} (FT) & 26.67 & 0.7378 & \textbf{4.2885}  & 0.1134
\\
FASRGAN (ours, FT) & \textbf{27.57} & \textbf{0.7809} & 5.0006 & \textbf{0.1063}
\\
Fs-SRGAN (ours, FT) & 25.82 & 0.7663 & \underline{4.9929}
 & \underline{0.1121}
 \\ 
 \hline

\end{tabular}
\vspace{1.5mm}
\end{center}
\end{table}

\begin{figure*}[]
	\setlength{\fsdurthree}{-0.4cm}
	\scriptsize
	\centering
	\begin{tabular}{c}
		\hspace{-0.4cm}
		\begin{adjustbox}{valign=t}
		\tiny
			\begin{tabular}{ccccccc}
			    \centering	\includegraphics[width=0.147\textwidth]{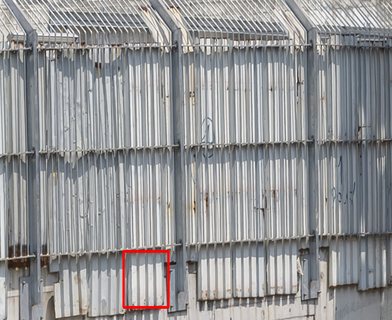} \hspace{\fsdurthree} &
				\includegraphics[width=\widthscalefour \textwidth]{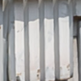} \hspace{\fsdurthree} &
				\includegraphics[width=\widthscalefour \textwidth]{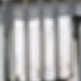} \hspace{\fsdurthree} &
				\includegraphics[width=\widthscalefour \textwidth]{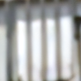} \hspace{\fsdurthree} &
				\includegraphics[width=\widthscalefour \textwidth]{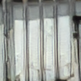}
				\hspace{\fsdurthree} &
				\includegraphics[width=\widthscalefour \textwidth]{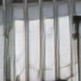}
				\hspace{\fsdurthree} &
				\includegraphics[width=\widthscalefour \textwidth]{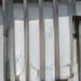}
				\\
				Canon \hspace{\fsdurthree} &
				HR \hspace{\fsdurthree} &
				ZSSR~\cite{shocher2018zero} 
				\hspace{\fsdurthree} &
				FSSR~\cite{fritsche2019frequency} 
				\hspace{\fsdurthree} &ESRGAN~\cite{wang2018esrgan} (FT) \hspace{\fsdurthree} &
				FASRGAN (ours, FT)
				\hspace{\fsdurthree} &
				Fs-SRGAN (ours, FT)
				\\
				016 \hspace{\fsdurthree} &
				PSNR/PI/LPIPS \hspace{\fsdurthree} &
				22.57/7.1428/0.2866 \hspace{\fsdurthree} &
				22.15/7.8434/0.3192 \hspace{\fsdurthree} &
				21.58/5.5788/0.1274 \hspace{\fsdurthree} &
				22.99/5.9955/0.1065 \hspace{\fsdurthree} &
				21.48/6.0169/0.1064 
			\end{tabular}
		\end{adjustbox}
		\vspace{0.5mm}
		\\
		\hspace{-0.4cm}
		\begin{adjustbox}{valign=t}
		\tiny
			\begin{tabular}{ccccccc}
			    \centering	\includegraphics[width=0.147\textwidth]{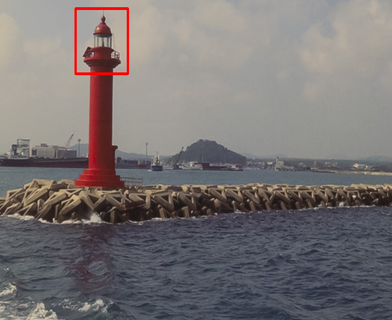} \hspace{\fsdurthree} &	\includegraphics[width=\widthscalefour \textwidth]{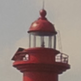} \hspace{\fsdurthree} &	\includegraphics[width=\widthscalefour \textwidth]{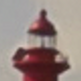} \hspace{\fsdurthree} &	\includegraphics[width=\widthscalefour \textwidth]{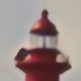} \hspace{\fsdurthree} &	\includegraphics[width=\widthscalefour \textwidth]{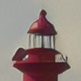}
				\hspace{\fsdurthree} &	\includegraphics[width=\widthscalefour \textwidth]{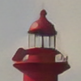}
				\hspace{\fsdurthree} &	\includegraphics[width=\widthscalefour \textwidth]{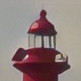}
				\\
				Canon \hspace{\fsdurthree} &
				HR \hspace{\fsdurthree} &
				ZSSR~\cite{shocher2018zero} 
				\hspace{\fsdurthree} &
				FSSR~\cite{fritsche2019frequency} 
				\hspace{\fsdurthree} &
				ESRGAN~\cite{wang2018esrgan} (FT) \hspace{\fsdurthree} &
				FASRGAN (ours, FT)
				\hspace{\fsdurthree} &
				Fs-SRGAN (ours, FT)
				\\
				030 \hspace{\fsdurthree} &
				PSNR/PI/LPIPS \hspace{\fsdurthree} &
				31.50/7.9916/0.1578 \hspace{\fsdurthree} &
				29.66/7.4778/0.1520 \hspace{\fsdurthree} &
				28.93/3.1849/0.0816 \hspace{\fsdurthree} &
				30.99/4.2431/0.0768 \hspace{\fsdurthree} &
				28.13/4.5620/0.0723 
			\end{tabular}
		\end{adjustbox}
		\vspace{0.5mm}

		\\
		\hspace{-0.4cm}
		\begin{adjustbox}{valign=t}
		\tiny
			\begin{tabular}{ccccccc}
			    \centering	\includegraphics[width=0.147\textwidth]{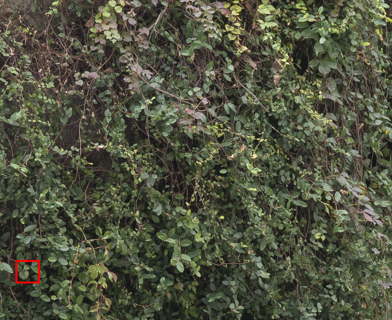}\hspace{\fsdurthree} &	\includegraphics[width=\widthscalefour \textwidth]{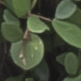} \hspace{\fsdurthree} &	\includegraphics[width=\widthscalefour \textwidth]{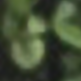} \hspace{\fsdurthree} &	\includegraphics[width=\widthscalefour \textwidth]{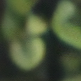} \hspace{\fsdurthree} &	\includegraphics[width=\widthscalefour \textwidth]{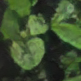}
				\hspace{\fsdurthree} &	\includegraphics[width=\widthscalefour \textwidth]{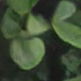}
				\hspace{\fsdurthree} &	\includegraphics[width=\widthscalefour \textwidth]{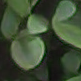}
				\\
				Nikon \hspace{\fsdurthree} &
				HR \hspace{\fsdurthree} &
				ZSSR~\cite{shocher2018zero} 
				\hspace{\fsdurthree} &
				FSSR~\cite{fritsche2019frequency} 
				\hspace{\fsdurthree} &
				ESRGAN~\cite{wang2018esrgan} (FT) \hspace{\fsdurthree} &
				FASRGAN (ours, FT)
				\hspace{\fsdurthree} &
				Fs-SRGAN (ours, FT)
				\\
				045 \hspace{\fsdurthree} &
				PSNR/PI/LPIPS \hspace{\fsdurthree} &
				23.32/8.1629/0.3535 \hspace{\fsdurthree} &
				22.80/8.3245/0.3514 \hspace{\fsdurthree} &
				21.04/3.7778/0.2069 \hspace{\fsdurthree} &
				22.71/4.3211/0.1688 \hspace{\fsdurthree} &
				22.34/3.9972/0.1530 
			\end{tabular}
		\end{adjustbox}
		\vspace{0.5mm}

 	\end{tabular}
	\caption{
		 The visual comparisons between our proposed methods and compared methods on RealSR(V3) test set for $4\times$. FT represents the model has been fine-tuned on RealSR (v3) training set.
	}
\label{fig:RealSR}
\end{figure*}

\subsection{Results on the Real-World Dataset}
We also benchmark our proposed methods on a publicly available real-world dataset to test the robustness. We adopt the test set from RealSR(V3)~\cite{cai2019toward} as the dataset and PSNR/SSIM/PI/LPIPS as evaluation metrics. As shown in the top part of Table~\ref{tab:realsr}, 
our FASRGAN and Fs-SRGAN obtain better PI and LPIPS than ESRGAN and comparable results with RankSRGAN, demonstrating that our proposed models have better robustness on real-world LR images. 

In addition, we used the training set from RealSR(V3) to fine-tune ESRGAN and our proposed methods. Both of them have run about 150k iterations, where the learning rate is initially set as $10^{-4}$ and decays a half every 50k iterations. We test the fine-tuned (FT) models on the test set, and also compare them with ZSSR~\cite{shocher2018zero} and the work from Fritsche \emph{et al.}~\cite{fritsche2019frequency} proposed for the AIM 2019 Challenge on Real World, denoted as FSSR. ZSSR is the first unsupervised SR method for non-ideal LR images. The codes and models of FSSR are publicly available, and we adopt the model TDSR of AIM for comparison. As shown in the bottom part of Table~\ref{tab:realsr}, our FASRGAN and Fs-SRGAN still obtain better results in LPIPS, indicating that our models are robust on the real-world images.

Visual results of the fine-tuned models and the compared methods are also presented in Fig.~\ref{fig:RealSR}. We can observe that the results generated by ZSSR and FSSR are heavily blurred, which brings a bad visual effect. The fine-tuned results from ESRGAN contain some artifacts and noises, resulting in unpleasing observation. While our fine-tuned FASRGAN and Fs-SRGAN reduce the artifacts and produce more pleasing results, demonstrating the robustness of our proposed models.

\section{Conclusion}
We propose two GAN-based models, FASRGAN and Fs-SRGAN, for SISR to overcome the limitations of existing methods. FASRAGN introduces a fine-grained attention mechanism into the GAN framework, where the discriminator has two outputs: a score for measuring the quality of the whole input and a fine-grained attention estimation for the input. The fine-grained attention delivers a fine-grained supervisor to the generator to ensure generation of pixel-wise photo-realistic images. The Fs-SRGAN shares the low-level feature extractor of the generator and the discriminator, reducing the number of parameters and improving the reconstruction performance. These two mechanisms are general and could be applied to other GAN-based SR models. Comparisons with other state-of-the-art methods on benchmark datasets demonstrate the effectiveness of our proposed methods.

\bstctlcite{IEEEexample:BSTcontrol}

\bibliographystyle{IEEEtran}
\bibliography{IEEEabrv,Main}

\ifCLASSOPTIONcaptionsoff
  \newpage
\fi

\end{document}